\documentclass[final,number, square, 3p,times]{elsarticle}

\usepackage[utf8]{inputenc}
\usepackage{fontenc}
\usepackage{lmodern}
\usepackage{amsmath,amssymb,amsthm,bm, amsfonts}
\usepackage{dsfont}
\usepackage{graphicx,color}
\usepackage{caption}
\usepackage{subcaption}
\usepackage{booktabs}
\usepackage[colorlinks]{hyperref}
\usepackage[dvipsnames]{xcolor}
\usepackage[capitalise,noabbrev]{cleveref}
\usepackage{float}
\usepackage[section]{placeins}
\usepackage{algpseudocode}
\usepackage{algorithm2e}
\usepackage{algorithmicx}
\usepackage{multirow}
\usepackage{parskip}
\usepackage{soul} 
\usepackage{bbm}
\usepackage{makecell}

\newcommand{\Temp}{T_\text{roll}}
\newcommand{\Roll}{R_\text{roll}}
\newcommand{\Dir}{D_\text{roll}}
\newcommand{\RD}{\text{RD}}
\newcommand{\TD}{\text{TD}}

\newcommand{\boldface}[1]{\boldsymbol{#1}}  %
\newcommand{\bfa}{\boldface{a}}

\newcommand{\bfc}{\boldface{c}}

\newcommand{\bfg}{\boldface{g}}

\newcommand{\bfi}{\boldface{i}}
\newcommand{\bfj}{\boldface{j}}
\newcommand{\bfk}{\boldface{k}}

\newcommand{\bfp}{\boldface{p}}
\newcommand{\bfq}{\boldface{q}}
\newcommand{\bfr}{\boldface{r}}

\newcommand{\bfu}{\boldface{u}}
\newcommand{\bfv}{\boldface{v}}

\newcommand{\bfx}{\boldface{x}}
\newcommand{\bfy}{\boldface{y}}
\newcommand{\bfz}{\boldface{z}}

\newcommand{\bfF}{\boldface{F}}
\newcommand{\bfG}{\boldface{G}}

\newcommand{\bfI}{\boldface{I}}

\newcommand{\bfQ}{\boldface{Q}}
\newcommand{\bfR}{\boldface{R}}

\newcommand{\bfT}{\boldface{T}}

\newcommand{\bfrho}{\boldsymbol{\rho}}

\newcommand{\calD}{\mathcal{D}}

\newcommand{\calF}{\mathcal{F}}
\newcommand{\calG}{\mathcal{G}}

\newcommand{\calK}{\mathcal{K}}

\newcommand{\calP}{\mathcal{P}}
\newcommand{\calQ}{\mathcal{Q}}
\newcommand{\calR}{\mathbb{R}}

\newcommand{\calT}{\mathcal{T}}

\newcommand{\partderiv}[2]{\frac{\partial #1}{\partial #2}}

\newcommand{\Rset}{\mathbb{R}}

\newlength{\boxwidth}
\def\dd{\;\!\mathrm{d}}

\def\btheorem{\begin{theorem}}
\def\etheorem{\end{theorem}}
\def\blemma{\begin{lemma}}
\def\elemma{\end{lemma}}
\def\bproposition{\begin{proposition}}
\def\eproposition{\end{proposition}}
\def\bcorollary{\begin{corollary}}
\def\ecorollary{\end{corollary}}
\def\bdefinition{\begin{definition}}
\def\edefinition{\end{definition}}
\def\bexample{\begin{example}}
\def\eexample{\end{example}}
\def\bremark{\begin{remark}}
\def\eremark{\end{remark}}

\def\fg{\boldsymbol}

\newcommand{\be}{\begin{equation}}
\newcommand{\ee}{\end{equation}}
\newcommand{\beq}{\begin{eqnarray}}
\newcommand{\eeq}{\end{eqnarray}}
\newcommand{\bem}{\begin{multline}}
\newcommand{\eem}{\end{multline}}
\newcommand{\ba}{\begin{align}}
\newcommand{\ea}{\end{align}}
\newcommand{\bfzero}{{\fg0}}

\newcommand{\argminWithArgs}{\operatornamewithlimits{arg\ min}}

\newcommand{\argmaxWithArgs}{\operatornamewithlimits{arg\ max}}

\newcommand{\SO}{\mathrm{SO}(3)}
\newcommand{\Si}{\mathrm{S}^1}
\newcommand{\Sii}{\mathrm{S}^2}

\newcommand{\Sn}{\mathrm{S}^n}

\journal{arXiv}

\begin{document}

\sffamily 

\begin{frontmatter}

\title{\textbf{TexFlow}: Learning Crystallographic Texture on SO(3) Manifold via Normalizing Flows}

\author[a]{Gowtham Nimmal Haribabu}
\author[a]{Tarakram Ramgopal}
\author[a]{Divakar Vetriveeran}
\author[a]{Siddhant Kumar}

\ead{Sid.Kumar@tudelft.nl}
\address[a]{Department of Materials Science and Engineering, Delft University of Technology, 2628CD Delft, The Netherlands}

\begin{abstract}
Crystallographic texture is an important microstructural feature that statistically describes the orientations of grains in polycrystalline materials. Conventional texture analysis relies on the orientation distribution functions and pole figures, which are essential for interpreting the texture of a given material specimen but do not, by themselves, provide a direct framework for predictive modeling of process-structure-property linkages. Data- and machine learning-based approaches offer a means to address this gap; however, a key challenge lies in developing representations and learning frameworks that account for the manifold structure of rotations. In particular, many existing machine learning techniques are designed for Euclidean data and therefore are not readily applicable to rotations in the non-Euclidean group $\SO$. Here we introduce \textbf{TexFlow}, a conditional normalizing flow framework for texture modeling that learns a probability distribution over rotations in $\SO$ by construction. Rotations are encoded in a six-dimensional Gram-Schmidt orthogonalization representation which is then mapped back to valid rotations $\bfR\in\SO$. The framework supports conditional probabilistic modeling of texture $p(\bfR|\bfc)$, where $\bfc$ may represent processing parameters like temperature or deformation state. To complement conventional texture-analysis visualizations, such as pole figures and sections of the orientation distribution function, we further introduce Hopf-fibration visualizations as a compact global representation of texture. TexFlow is validated on both unconditional and conditional experimental and simulated texture datasets spanning diverse material systems. We show that TexFlow can accurately capture multimodal textures and their conditional evolution, providing a data-driven framework for establishing process-texture linkages.
\end{abstract}


\end{frontmatter}

\section{Introduction}
\label{sec:Introduction}

Crystallographic texture describes the statistical distribution of grain orientations in polycrystalline materials \citep{Bunge1982TextureScience,Roe1965DescriptionInversion}. As texture is both influenced by processing and determines the properties of materials, it plays a critical role in bridging the Process-Structure-Property (PSP) linkages. Consequently, prediction of texture is an important challenge  for predicting material performance and optimizing process conditions for targeted properties \citep{Fullwood2010MicrostructureOptimization,Adams2013Microstructure-SensitiveOptimization}. 

Comprehensively mapping the PSP space using experimental data alone is generally impractical, as techniques such as X-ray diffraction and electron backscatter diffraction (EBSD) are time-consuming and expensive. Predictive models are therefore needed to efficiently fill gaps in the PSP space where experimental data are unavailable. Physics-based models, such as Taylor models \citep{Taylor1938PlasticMetals,Tutcuoglu2019High-Copper,Kumar2020ACopper}, viscoplastic self-consistent (VPSC) models \citep{Lebensohn1993AAlloys}, and crystal plasticity finite element methods (CPFEM) \citep{Peirce1983MaterialSolids,Roters2010OverviewApplications,Roters2019DAMASKScale,Yaghoobi2022PRISMS-PlasticityPipeline}, have been successfully used to predict texture evolution under a range of thermomechanical processes \citep{Bronkhorst1992PolycrystallineMetals,Anand1996APlasticity,Lee1996TextureCopper}.
However, these methods are computationally expensive, with simulation runtimes ranging from hours to weeks or even months for large-scale, multiscale systems. This limits their applicability for virtual high-throughput screening of PSP linkages, particularly in data-scarce regimes.  Additionally, from a process-design perspective, there is a need for models that enable high-throughput exploration of processing pathways, quantification of texture variability and uncertainty, and inverse-design workflows for optimizing processing parameters to achieve target texture states or material properties. These requirements motivate the development of machine learning (ML)-based probabilistic surrogate models for predicting texture as a function of process parameters.

ML models have already shown recent success for discovering PSP linkages \citep{McDowell2016TheInitiative,Choudhary2022RecentScience}. In the context of microstructure and texture, ML has been adopted as a data-driven tool for representation learning, prediction, and inverse design \citep{Schmidt2023APlasticity}. A notable work by Wanni et al. introduced the Texture Adaptive Clustering and Sampling (TACS) framework, which uses unsupervised learning to learn a reduced order representation for large EBSD datasets \citep{Wanni2024MachineRepresentation}. Hybrid models have been proposed to accelerate texture prediction, where CPFEM simulations are combined with neural networks and regression models to predict texture evolution \citep{Rezaei2025CrystallographicPlasticity} and a related work has linked texture descriptors to plastic anisotropy using deep learning models trained on large CPFEM datasets \citep{Park2023DevelopmentLinkage}. More recently, Buzzy \emph{et al.} demonstrated that generative models based on conditional normalizing flows can be used for active learning and inverse design of polycrystalline textures \citep{Buzzy2024ActiveFlows}.

Despite substantial progress in both classical and ML-based texture modeling, a key challenge remains in the choice of an appropriate representation of texture on which these models can operate. Classically, crystallographic texture has been represented as a probability density over the space of orientations, leading to the orientation distribution function (ODF) formalism \citep{Bunge1982TextureScience,Roe1965DescriptionInversion,Matthies1988SomeFigures}. These approaches express the ODF using generalized spherical harmonic expansions, which provide a rigorous mathematical foundation but require truncation at finite order, limiting the resolution of sharp or highly localized textures. Non-parametric kernel-based estimators avoid series truncation but introduce sensitivity to kernel bandwidth and discretization choices \citep{Hielscher2008AAlgorithm,Bachmann2010TextureToolbox}. While these tools remain indispensable for texture analysis of a given material specimen, they are not designed to directly fill gaps in the PSP space through interpolation or extrapolation where texture data are unavailable.

Many existing ML approaches fill these PSP gaps by learning, as a function of process or property conditions, either the coefficients of generalized spherical harmonic expansions \cite{Buzzy2024ActiveFlows,Safi2026,Buzzy2026}, implicit generative models directly in Euler-angle coordinates \cite{Wanni2024MachineRepresentation}, or inverse pole figures using image-based convolutional architectures \cite{Iraki2024AccurateStatistics,Fernandez-Zelaia2024,Kaushik2022}. While these representations are convenient and physically interpretable, they are not sufficient to uniquely characterize the joint distribution on the rotation group $\SO$, as recently emphasized by \citet{Schaeben2025CommentRepresentation} in a critical assessment of commonly used texture representations. This is because these representations do not preserve the topology of $\SO$. For example, Euler-angle representations introduce coordinate singularities and discontinuities, which can make otherwise nearby rotations appear artificially distant in the representation space. Pole figures map the $\SO$ space onto the two-dimensional sphere $\Sii$, resulting in a loss of information. In \autoref{subsec:rot_rep} and \autoref{subsec:prob_rot}, we provide a detailed discussion of different rotation representations and probabilistic modeling approaches, along with their respective limitations. These limitations can adversely affect the ability of ML models to learn smooth and physically meaningful mappings between processing conditions and texture.

This motivates the development of an ML approach for texture that respects the topology of rotations in $\SO$ and adopts a generative framework to learn the conditional probability distribution of rotations given the processing conditions. Analogous approaches in computer graphics and computer vision have recently attracted interest, with generative models being developed to capture complex, multimodal distributions on non-Euclidean manifolds \citep{Liu2023DelvingModeling,Zhou2018OnNetworks,pmlr-v119-rezende20a,Murphy2021Implicit-PDF:Manifold}.

\begin{figure}[t]
    \centering
    \includegraphics[width=\textwidth]{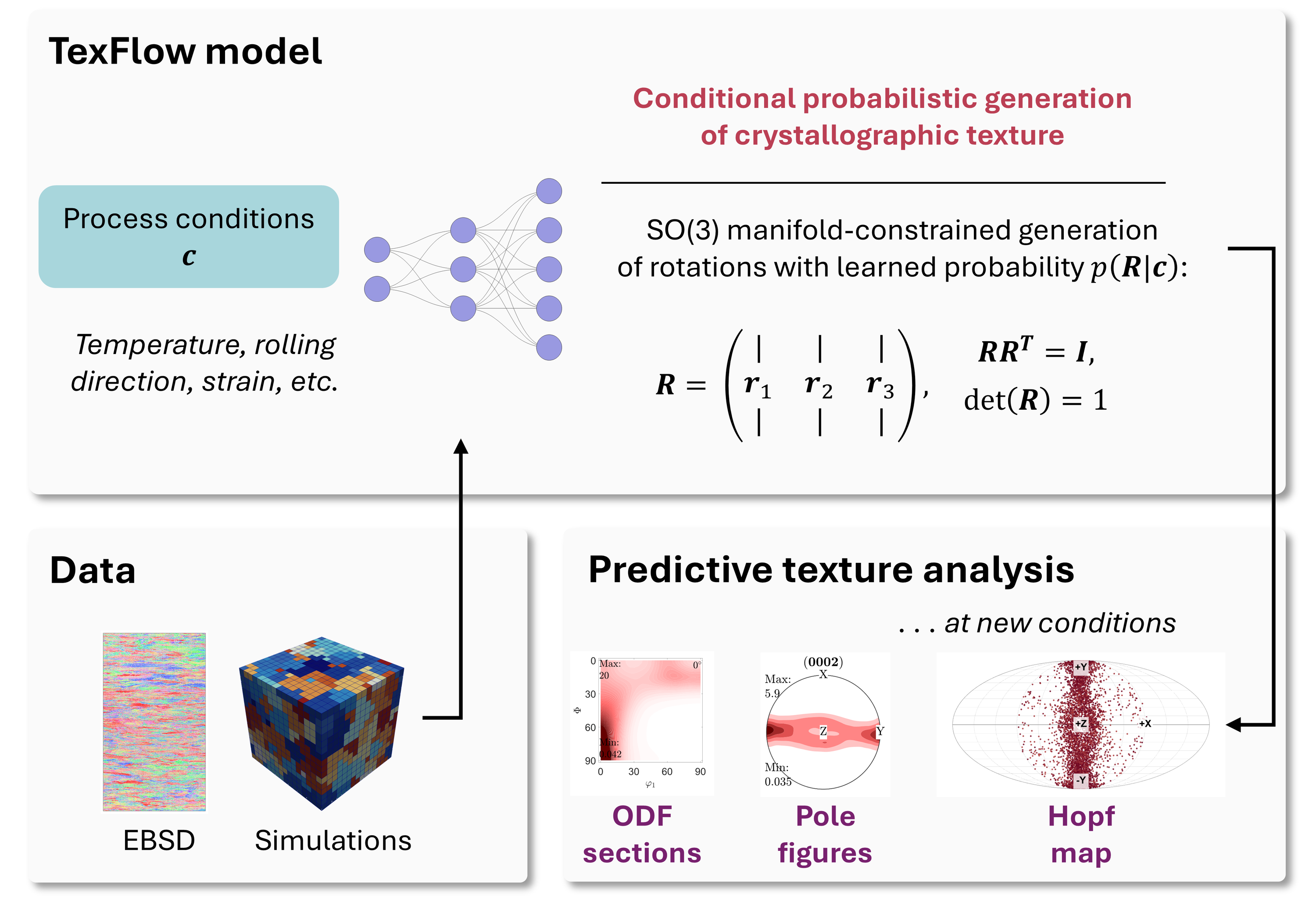}
    \caption{
Overview of the proposed \textbf{TexFlow} framework for conditional probabilistic generation of crystallographic texture. The ML model (described in detail later) takes the processing conditions $\bfc$, such as temperature, rolling direction, and simulation time, as input and probabilistically generates rotations $\bfR \in \SO$ according to the learned conditional distribution $p(\bfR|\bfc)$. The model is trained to match this distribution to the ground-truth distribution of rotations obtained from experiments (e.g., EBSD) or simulations (e.g., CPFEM) under the corresponding processing conditions $\bfc$. Once trained, the learned conditional distribution can be sampled to predict texture at previously unseen processing conditions. The generated textures can then be analyzed using conventional visualization tools, such as ODF sections and pole figures, or the Hopf-fibration maps introduced in this work. 
}
    \label{fig:overview}
\end{figure}

Here, we introduce \textbf{TexFlow}, a conditional normalizing flow framework for probabilistic modeling of crystallographic texture  that respects the topology of rotation in $\SO$. \autoref{fig:overview} provides an overview of the framework. Orientations are represented using a six-dimensional Gram-Schmidt orthogonalization (GSO) encoding, enabling unconstrained learning while preserving valid rotations by construction. Processing parameters ($\bfc$) are incorporated explicitly as conditioning variables, allowing the model to learn texture as conditional distributions $p(\bfR|\bfc)$ for $\bfR\in\SO$. Beyond density estimation and sampling, we employ Hopf fibration as a complementary visualization tool for analyzing orientation distributions, providing a compact and globally consistent view of texture. The framework is validated on simulation and experimental datasets, demonstrating accurate reconstruction of multimodal texture and conditional evolution. Complementing expensive simulations and experiments, the proposed framework enables efficient exploration of process-texture linkages.

The manuscript is organized as follows. \autoref{subsec:rot_rep} and \autoref{subsec:prob_rot} provide an exposition of rotation representations and probability distributions, respectively, motivating the design choices adopted in this work. \autoref{subsec:texflow_desc} describes the methodology of the TexFlow framework. \autoref{sec:results} presents a series of benchmarks evaluating the performance of TexFlow on simulated and experimental texture datasets, followed by the conclusions in \autoref{sec:conclusion}.

\section{Rotation parameterizations: mathematical preliminaries and limitations}
\label{subsec:rot_rep}

The orientation of a crystal lattice inside a three-dimensional polycrystalline material is mathematically expressed by a rotation that maps the crystal reference frame to the sample reference frame. The set of all proper rotations forms the \emph{Special Orthogonal Group}: 
\be
\label{SO3}
    \SO = \{ \bfR \in \calR^{3\times3} : \bfR^\intercal\bfR = \bfI,\ \det(\bfR)= +1 \}.
\ee
We also denote the $n$-sphere, i.e., the sphere of unit radius in $(n+1)$-dimensional Euclidean space, as
\be
\Sn = \{ \bfx\in\Rset^{n+1} : |\bfx|=1 \}.
\ee
This notation will be used throughout the subsequent discussion.

Although $\bfR\in\SO$ is a $3\times 3$ matrix, the set $\SO$ forms a smooth three-dimensional manifold. A rotation $\bfR\in\SO$ can be parameterized in several equivalent ways that are widely used in crystallography, mechanics, and robotics. However, each parameterization has some trade-offs. Here, we briefly review the most common parameterizations and motivate the choice of an appropriate representation for the generative modeling task at hand. Refs.~\citep{Hashim2019SpecialChallenges, Makinen2008RotationVectors} provide detailed derivations and formulas for converting between various parameterizations.

As a prologue, when parameterizing a manifold (the rotation manifold in this case), the dimensionality of the parameterization plays an important role. If the parameterization is too low-dimensional, it may exhibit singularities and discontinuities. Two rotations that are close on the $\SO$ manifold may be far apart in the chosen parameterization, creating discontinuous loss landscapes and unstable gradients during learning.  On the other hand, the parameterization can also be too high-dimensional and overparameterized. In this case, singularities and discontinuities disappear, but additional constraints arise to ensure that the representation lies on the manifold. However, even with sufficient training, the output of a generative model may fail to satisfy these constraints and therefore may not lie on the $\SO$ manifold, resulting in invalid rotations.

Euler's rotation theorem states that any rotation in three-dimensional space can be represented as a single rotation about an axis. Consequently, three is the minimum number of dimensions required to parameterize a rotation: one for the angle of rotation and two for the orientation of the rotation axis. However, three-dimensional parameterizations of rotations, also known as minimal parameterizations, suffer from discontinuities. In fact, \citet{Stuelpnagel1964} proved that it is topologically impossible to construct a three-dimensional parameterization of the rotation group without introducing singularities. \citet{Hopf1964} showed that \textbf{a minimum of five dimensions is required for a parameterization to provide a continuous representation of the rotation group}. We quote the succinct proof by \citet{Zhou2018OnNetworks} for the topology-inclined readers, demonstrating why at least five dimensions are required to parameterize the rotation group without discontinuities. 

\textit{``The 3D rotation group $\SO$ is homeomorphic to the real projective space $\Rset P^3$. The space $\Rset P^n$ is defined as the quotient space of $\Rset^{n+1}\backslash \{0\}$ under the equivalence relation that $x\sim\lambda x$ for all $\lambda \neq 0$. $\dots$ 
Based on standard embedding and non-embedding results in topology \cite{Davis1998EmbeddingsOR}, we know that $\Rset P^3$ (and thus $\SO$) embeds in $\Rset^5$ with the Euclidean topology, but does not embed in $\Rset^d$ for any $d<5$. By the definition of embedding,
there is no homeomorphism from $\SO$ to any subset of
$\Rset^d$ for any $d<5$, but a continuous representation requires
this. Thus, there is no such continuous representation [for $d<5$].'' }

\textbf{Euler angles:} In crystallographic texture analysis, proper Euler angles are commonly written as, e.g.,  $(\varphi_1, \Phi, \varphi_2)$ in the Bunge $Z\!X\!Z$ convention. The crystal orientation is defined by a rotation about the original $z$-axis by angle $ \varphi_1 $, followed by a rotation about the intermediate $x$'-axis by $\Phi$, and finally by a rotation about the final $z''$-axis by $ \varphi_2 $.  In an active column-vector convention, the rotation matrix corresponding to $(\varphi_1, \Phi, \varphi_2)$ is given by
\be
\label{eq:euler_define}
    {\bfR}( \varphi_1, \Phi, \varphi_2) = {\bfR}_z( \varphi_1) {\bfR}_x(\Phi) {\bfR}_z(\varphi_2),
    \qquad \text{with}\quad 
    (\varphi_1, \Phi, \varphi_2) \in [0,2\pi)\times[0,\pi]\times[0,2\pi),
\ee
where $\bfR_{(\cdot)}$ represents the constituent rotation around the $(\cdot)$-axis.
Euler angles are intuitive and form the basis for classical texture analysis, but they suffer from issues at $\Phi=0$ and $\Phi=\pi$, where the first and third rotation axes become aligned. Consequently, changes in $\varphi_1$ and $\varphi_2$ are no longer independently distinguishable, giving rise to the coordinate singularity commonly known as gimbal lock \cite{gimbal_lock_wiki}. Additionally, $\varphi_1$ exhibits topological discontinuities due to periodicity, as distinct values $\varphi_1 = 0$ and $\varphi_1 \rightarrow 2\pi$ (approached as a limit) represent the same rotation, and likewise for $\varphi_2$. Previous works \cite{SaxenaEtAl2009,Zhou2018OnNetworks} have shown that Euler angles are not suitable parameterizations for predicting or generating rotation due to the aforementioned limitations.

\textbf{Angle-axis representation:}  This representation parametrizes any rotation as a turn by an angle $\alpha\in[0,\pi]$ around a fixed unit direction ${\bfu} \in \Sii$ (i.e., axis of rotation). Its map to $\SO$ is given by the Rodrigues' formula: 
\be
\label{eq:axis_angle_short}
    \bfR(\alpha,{\bfu})
    =
    \exp(\alpha{[\bfu]_\times})
    =
    \bfI+(\sin\alpha){[\bfu]_\times}
    +(1-\cos\alpha){[\bfu]_\times^2},
    \qquad\text{with}\quad
    [\bfu]_\times=\begin{bmatrix}
     0 & -u_3 & u_2 \\
     u_3& 0  & -u_1\\
     -u_2 & u_1 & 0
    \end{bmatrix}
\ee
for $\bfu=(u_1,u_2,u_3)^\intercal$ and $\|\bfu\|=1$. The angle-axis representation exhibits a singularity at $\alpha = 0$ where the rotation axis is undefined. Additionally, infinitely many arbitrary axes  yield the same identity rotation. Furthermore, a rotation of $\alpha =\pi$ around an arbitrary axis $\bfu$ is the same as  that around the axis $-\bfu$, which introduces a topological discontinuity in the parameterization of $\SO$.

\textbf{Minimal vectorial parameterizations:}  The rotation is parameterized as a vector $\bfrho\in\Rset^3$ , with several different possible formulations \cite{Bauchau2003} such as:
\begin{itemize}
    \item Rotation vector: $\bfrho = \alpha \bfu$
    \item Cayley–Gibbs–Rodrigues vector: $\bfrho = 2\tan\left(\alpha/2\right) \bfu$,
    \item Wiener–Milenkovic vector: $\bfrho = 4\tan\left(\alpha/4\right) \bfu$,
    \item Euler–Rodrigues vector: $\bfrho = 2\sin\left(\alpha/2\right) \bfu$,
\end{itemize}
where $\alpha\in\Rset$ and $\bfu\in\Sii$ denote the angle and axis of rotation, respectively. The rotation matrix $\bfR(\bfrho)$ can be obtained via \eqref{eq:axis_angle_short} by substituting out $\bfu$ in terms of $\bfrho$. 

The rotation vector exhibits topological discontinuities, as concentric shells centered at the origin and $2\pi$ radial distance apart, i.e., $\|\bfrho\|+2n\pi$ for any integer $n$, map to the same rotation in $\SO$. Additionally, antipodal points $\bfrho=\pi\bfu$ and $\bfrho=-\pi\bfu$  map to the same rotation but are discontinuous.

The other vectorial representations, which are based on trigonometric functions of the rotation angle, exhibit singularities and discontinuities similar to the aforementioned ones. Tangent-based representations further introduce additional singularities at specific rotation angles; e.g., at $\alpha = n\pi$ for the Cayley–Gibbs-Rodrigues vector and at $\alpha = 2n\pi$ for the Wiener-Milenkovic vector, where $n$ is any integer. These singularities can be avoided by restricting $\alpha$ to an appropriate range; however, this requires the use of shadow parameters that introduce discontinuities \cite{Bauchau2003,schaub1996stereographic}.

\textbf{Unit quaternions:} A quaternion is an extension of complex numbers to four components and of the form:
\be
\bfq = q_0 + q_1 \hat \bfi + q_2 \hat \bfj + q_3 \hat \bfk
\ee
with non-commutative multiplication, i.e., $\bfp\bfq\neq\bfq\bfp$ and $\hat \bfi^2 = \hat \bfj^2 = \hat \bfk^2 = \hat \bfi\hat \bfj\hat \bfk=-1$. A unit quaternion, i.e., $\|\bfq\|^2=1$, represents a rotation from $\SO$ \cite{Man2022CrystallographicRepresentations}. A Euclidean vector $\bfz\in\Rset^3$ can be represented as a pure quaternion (i.e., quaternion with zero real component) of the form $\bfz= (z_1,z_2,z_3) = z_1 \hat \bfi + z_2 \hat \bfj + z_3 \hat \bfk$ and rotated by a unit quaternion $\bfq$ under the operation
\be\label{eq:q_mult_rot}
\bfz' = \bfq \bfz \bfq^{-1},
\ee
where $\bfz'$ is a pure quaternion representing the rotated vector. The Euler–Rodrigues formula \cite{Siminovitch1997} connects unit quaternions with angle-axis representation as
\be
\bfq = \cos\left(\frac{\alpha}{2}\right) + \sin\left(\frac{\alpha}{2}\right)(u_1 \hat \bfi + u_2 \hat \bfj + u_3 \hat \bfk).
\ee

Unit quaternions avoid the classical gimbal lock problem associated with Euler angles. However, since unit quaternions provide a four-dimensional parameterization, and fewer than five dimensions, they are discontinuous as a parameterization of $\SO$ rotations (see Appendix D of ref.~\cite{Zhou2018OnNetworks} for a proof specific to unit quaternions). This can manifest in learning problems in several ways. For example, the antipodes $\bfq$ and $-\bfq$ represent the same rotation (see \eqref{eq:q_mult_rot}); therefore, unit quaternions provide a double cover of $\SO$ \cite{Man2022CrystallographicRepresentations}. This introduces a topological discontinuity in the parameterization of rotations and can make the loss landscape of optimization or model training discontinuous \cite{Aiersilan2025}. Previous works \cite{SaxenaEtAl2009,Zhou2018OnNetworks} have shown that these discontinuities can hinder the learning of rotations with unit quaternions. Additionally, in optimization or generative tasks where unit quaternions are used as state variables, the unit-norm constraint may not be satisfied  \cite{Diebel}, in which case the corresponding operations do not represent rotations in $\SO$.

\textbf{Rotation matrix:} One may directly use the $3\times 3$ rotation matrix $\bfR$ directly to represent and generate the rotations. However, directly generating the nine entries of $\bfR$ can violate the orthogonality constraint: $\bfR^\intercal\bfR = \bfI$ and determinant constraint: $\det(\bfR)=+1$, thereby producing matrices that are invalid rotations. Therefore, a rotation parametrization used in gradient descent based learning should either enforce these constraints by construction or include a projection back onto $\SO$.

There are several other parameterizations, such as the Cayley–Klein parameters (four complex dimensions, or eight real dimensions), the direction cosine matrix (nine dimensions), and the Tait–Bryan angles (three dimensions). However, all of these parameterizations exhibit one or more of the aforementioned issues.

\section{Probability on rotations: mathematical preliminaries and limitations}
\label{subsec:prob_rot}

Crystallographic texture is essentially a probability distribution over $\SO$, i.e., the space of lattice rotations, commonly referred to as the orientation distribution function (ODF). Let $p:\SO\times\Rset^{k}\rightarrow\Rset_{\geq 0}$ denote the ODF describing the conditional probability density of observing a grain with rotation $\bfR\in\SO$, conditioned on input variables $\bfc\in\Rset^k$, such as process parameters or target properties. The conditioning variables can be omitted to give the unconditional ODF, $p(\bfR)$; however, we retain the conditioning to present a general formulation. The ODF obeys
\be
\label{eq:odf_normal}
    \int_{\SO} p(\bfR|\bfc)\,\dd\mu(\bfR) = 1 \qquad \forall\ \bfc\in\Rset^k,
\ee
where $\dd\mu(\bfR)$ is the normalized Haar measure 
or the invariant volume of $\SO$ (analogous to $\dd x\dd y\dd z$ for Euclidean volume). For example, in Bunge Euler angles $(\varphi_1,\Phi,\varphi_2)$, the normalized Haar measure is $\dd\mu(\bfR) = \frac{1}{8\pi^2}\sin\Phi\,d\varphi_1\,d\Phi\,d\varphi_2$. A uniformly random texture corresponds to $p(\bfR|\cdot)=1$ with respect to $\dd\mu$, and texture intensity is commonly reported as multiples of this random distribution.

In a generative modeling task, the objective is either to learn $p(\bfR|\bfc)$ directly or to generate samples of rotations that implicitly follow the underlying conditional probability defined by $p(\bfR|\bfc)$.
Here, we briefly review existing techniques to motivate our methodological choice for learning such probability distributions.

\textbf{Generalized spherical harmonics:} A classical route to represent the ODF is to expand it in an orthonormal basis on $\SO$. The natural harmonic basis on $\SO$ is given by the Wigner $D$-functions, also referred to in texture analysis as generalized spherical harmonics (GSH). The ODF can be written as 
\be
\label{eq:gsh_odf}
p(\bfR|\bfc) = \sum_{\ell=0}^{\infty} \sum_{m=-\ell}^{\ell} \sum_{n=-\ell}^{\ell} c^{(\ell)}_{mn}(\bfc)\, D^{\ell}_{mn}(\bfR),
\ee
where $D^{\ell}_{mn}(\bfR)$ are the matrix elements of the irreducible representations of $\SO$, and $c^{(\ell)}_{mn}$ are texture coefficients. The GSH coefficients could be used to build process-texture or texture-property linkages. Each texture is represented by a finite set of GSH coefficients and then a regression model can learn the relationship between processing parameters and the coefficients:
\be
\label{eq:gsh_psp_regression}
    \bfc \longmapsto \left\{c^{(\ell)}_{mn}(\bfc)\right\}_{0\leq \ell\leq L}.
\ee

This approach is attractive because it connects to classical texture analysis. However, there are certain limitations. Firstly, a finite cutoff $L$ restricts the resolution of the reconstructed ODF and may smooth out sharp peaks. The truncated harmonic expansion can generate oscillatory or non-positive artifacts (similar to Gibbs phenomenon in Fourier series), especially near localized texture components. Secondly, the GSH functions are defined in terms of Euler angles and suffer from discontinuities and singularities (and likewise for GSH variants based on other representations such as quaternions), due to the fact that they provide fewer than five-dimensional parameterizations of rotations (see discussion in \autoref{subsec:rot_rep}). Finally, rather than learning the orientation distribution directly as a probabilistic model from which samples can be drawn, regression on individual coefficients treats the spectral parameterization as a discriminative task, which may struggle to generalize in sparse data regimes.

\textbf{Kernel-based ODF:} Another common technique for estimating the ODF is to place a rotational kernel $\calK:\SO\rightarrow\mathbb{R}_{\geq0}$ (e.g., von-Mises–Fisher or de La Vallee Poussin kernel) centered at each measured rotation in $\{\bfR_i\in\SO : i=1,\dots,N\}$ and define the probability distribution as:
\be
\label{eq:kde_odf}
   p(\bfR) = \frac{1}{N} \sum_{i=1}^{N} \calK(\bfR\bfR_i^{-1}), 
   \qquad
   \text{with}\quad
   \int_{\SO} \calK(\bfR)\dd\mu(\bfR)=1.
\ee
Kernel methods can approximate an unconditional ODF, but their nonparametric nature does not admit conditional modeling wherein the ODF is expressed as a function of process or property conditions.

\textbf{Pole figures and image-based approaches:} One can also project a rotation $\bfR$ to lower dimensions by considering the rotated orientation $\bfR\bfu$ of a chosen pole or reference direction $\bfu \in \Sii$ of the reference lattice. The texture can then be analyzed as a probability distribution of the pole orientation, i.e., $p(\bfR\bfu)$. However, as a low-dimensional projection of $\bfR$, this representation loses information encoded in $p(\bfR)$, and it may not be possible to recover the latter from the former. Furthermore, it is common to generate heatmaps of pole orientation data, such as pole figures and inverse pole figures, using stereographic projections, which further obscure key texture information originally present in $\SO$ data. Ref. \cite{Schaeben2025CommentRepresentation} and our results in \autoref{subsec:diagnostic} show examples in which two very different textures yield identical pole figures. This ambiguity and loss of information would also propagate to downstream machine learning tasks that use image-based approaches, such as convolutional neural networks operating on pole figures as images \cite{Iraki2024AccurateStatistics}.

\section{TexFlow: methodology}
\label{subsec:texflow_desc}

In light of the limitations of existing rotation parameterizations (Euler angles, quaternions, rotation matrices, etc.) and rotation probability formulations (GSH, kernel methods, pole figures, etc.), we propose TexFlow, a framework for texture analysis that
\begin{itemize}
\item operates on a \textbf{six-dimensional Gram-Schmidt orthogonal (GSO) parameterization} \citep{Zhou2018OnNetworks,Geist2024LearningSO3,Levinson2020AnEstimation,Bregier2021DeepStudy} to  \ul{identically satisfy continuity and manifold structure} in the rotation representation and the resulting probability distribution;
\item uses a \textbf{normalizing flow} (NF) \cite{Dinh2014NICE:Estimation, Dinh2016DensityNVP,Papamakarios2017} generative architecture to model \ul{texture as a conditional probability distribution $p(\bfR|\bfc)$ on the $\SO$ manifold directly}, without any loss of information from projection to lower dimensions.
\end{itemize}
Note that the combination of the GSO parameterization and NF is essential, as vanilla NF operating in Euclidean space does not respect the $\SO$ manifold structure. In the following, we provide a detailed explanation of the GSO parameterization and its integration with NFs.

\subsection{Gram-Schmidt orthogonal parameterization of rotation} 

\begin{figure}[H]
    \centering
    \includegraphics[width=\textwidth]{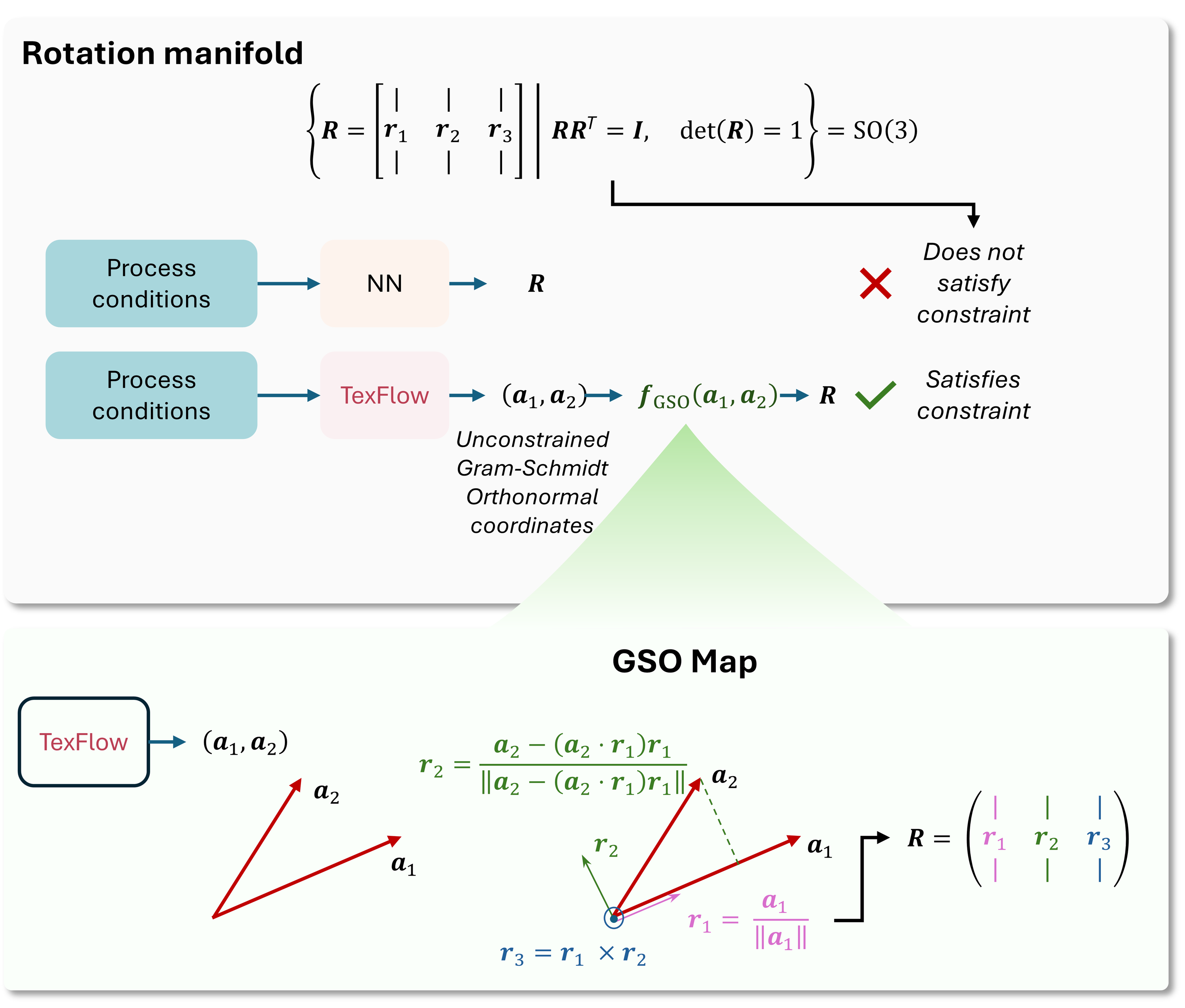}
    \caption{ 
    \textbf{Schematic illustration of the Gram-Schmidt Orthogonal (GSO) parametrization used in TexFlow.} A conventional neural network may produce a matrix that does not satisfy the constraints of the rotation manifold. In TexFlow, the process conditions are provided to the flow model, which predicts an unconstrained pair of vectors. The GSO map $f_\text{GSO}$ takes the unconstrained vectors as input and outputs the valid rotation matrix. The lower panel illustrates the normalization, orthogonal projection and cross-product operations used to obtain the rotation matrix while satisfying the manifold constraints.
    }
    \label{fig:gso_map}
\end{figure}

Let $\calG_\text{SO(3)}= \{(\bfa_1,\bfa_2):\bfa_1,\bfa_2\in\Rset^3\backslash\{\bm{0}\},|\bfa_1\cdot\bfa_2|\neq 1\}$ denote the set of pairs of unconstrained, non-parallel, non-zero vectors. We then use $(\bfa_1,\bfa_2)\in\calG_\text{SO(3)}$ as a six-dimensional unconditional parameterization of $\SO$. We apply the Gram–Schmidt process ($f_\text{GSO}$) to obtain three orthonormal vectors, $\bfr_1$, $\bfr_2$, and $\bfr_3$, which constitute the columns of $\bfR \in \SO$, as follows:
\be \label{eq:gso_to_r_map}
\begin{aligned}
\calG_\text{SO(3)}\rightarrow \SO: \quad &f_\text{GSO}(\bfa_1,\bfa_2) = \bfR =  \begin{bmatrix}
    | & | & |\\
    \bfr_1 & \bfr_2 & \bfr_3\\
    | & | & |  \\
\end{bmatrix}
\qquad \\
&
\text{with}
\quad
\bfr_1 = \frac{\bfa_1}{\|\bfa_1\|},
\quad 
\bfr_2 = \frac{\bfa_2-(\bfa_2\cdot\bfr_1)\bfr_1}{\|\bfa_2-(\bfa_2\cdot\bfr_1)\bfr_1\|}, 
\quad 
\bfr_3 = \bfr_1 \times \bfr_2.
\end{aligned}
\ee
Essentially, the map \textit{(i)} first normalizes $\bfa_1$ and assigns it to the first column, $\bfr_1$; \textit{(ii)} projects $\bfa_2$ onto $\bfr_1$ and subtracts the projection from $\bfa_2$ to obtain a vector orthogonal to the first column, which is then normalized and assigned to the second column, $\bfr_2$; and \textit{(iii)} computes the third column, $\bfr_3$, as the cross product of the first two columns. Since the columns of $\bfR$ are orthonormal vectors by construction, $\bfR$ is automatically special orthogonal by definition and identically satisfies the manifold constraints: $\bfR^\intercal\bfR=\bfI$ and $\det(\bfR)=+1$. The GSO map is visualized in \autoref{fig:gso_map}.

Note that despite the GSO parameterization of $\SO$ being many-to-one, i.e., not bijective, it is continuous and free from singularities; see ref.~\cite{Zhou2018OnNetworks} for a detailed proof. Additionally, since $\bfa_1$ and $\bfa_2$ are unconstrained vectors in $\mathbb{R}^3$, the GSO parameterization is well suited for generative modeling of textures without requiring explicit enforcement of rotation manifold constraints.

To bypass the lack of bijectivity, we define the following map \cite{Zhou2018OnNetworks} from $\SO$ to GSO parameters by simply dropping the last column of the rotation matrix:
\be \label{eq:r_to_gso_map}
\SO\rightarrow\calG_\text{SO(3)} : \quad g_\text{GSO}\left( \begin{bmatrix}
    | & | & |\\
    \bfr_1 & \bfr_2 & \bfr_3\\
    | & | & |  \\
\end{bmatrix} \right)
= (\bfr_1,\bfr_2).
\ee
The map $g_\text{GSO}$ satisfies the identity:
\be
f_\text{GSO}\circ g_\text{GSO}(\bfR) = \bfR.
\ee

\subsection{Conditional normalizing flows on GSO representation of rotations} 

Normalizing flows (NFs) \cite{Kobyzev2019NormalizingMethods,Dinh2014NICE:Estimation,Dinh2016DensityNVP,Papamakarios2019NormalizingInference} are a class of generative models that learn complex, multimodal probability distributions by transforming them into a simple base distribution (e.g., a standard Gaussian distribution) through a sequence of trainable, bijective, and differentiable transformations. In conditional normalizing flows, these transformations are parameterized by neural networks that take the conditioning variables as input. During sampling, a sample is drawn from the base distribution and mapped through the learned transformations, conditioned on the input variables, to generate a sample from the target conditional distribution. 

In the context of this work, let $\bfz\in\Rset^6$ denote a latent space with a simple base distribution $p(\bfz)$. Here, we  choose a centered Beta distribution for its flexibility in achieving diverse distribution types~\cite{Crenshaw2024ProbabilisticFlows}:
\be
\label{eq:latent_beta}
p_z(\bfz) = 2B\left(\text{Beta}(\bfz|\alpha,\beta)-\frac{1}{2}\right),
\ee

where $\alpha, \beta>0$ are trainable parameters and $B>0$ is a hyperparameter.  Let $\calT_\theta:\calG_{\SO}\times\Rset^k\rightarrow\Rset^6$
denote a trainable bijective function---also called a flow model---from GSO representation of rotation space to the latent space (with inverse $\calT^{-1}_\theta:\Rset^6\times\Rset^k\rightarrow\calG_{\SO}$) such that
\be
\begin{aligned}
    \text{Normalizing direction:}\quad 
    &\bfz = \calT_\theta(\bfg;\bfc)
    &\text{with}\quad
    &\bfg=g_\text{GSO}(\bfR);\\
    \text{Sampling direction:}\quad
    &\bfR=f_\text{GSO}(\bfg)
    &\text{with}\quad
    &\bfg =\calT^{-1}_\theta(\bfz;\bfc).
\end{aligned}
\ee
Here, $\theta$ is the set of trainable parameters in $\calT$. Partial bijectivity is assumed in map from $\bfg$ to $\bfz$; the condition $\bfc\in\Rset^k$ is not transformed by the flow at any stage. The first equation is referred to as the \emph{normalizing} direction because it maps the data distribution of rotations to the simple base distribution $p(\bfz)$. The second equation is referred to as the \emph{sampling} direction because it transforms a sample drawn from the base distribution into the rotation space. For unconditional settings, the conditions $\bfc$ can simply be omitted from the architecture or set to null values.

On the output of $\calT^{-1}_\theta$ (equivalently, input of $\calT_\theta$), we relax the definition of $\calG_{\SO}$ to $\Rset^6$ and use the two interchangeably throughout the remainder of this discussion. This is justifiable because a pair of  $\Rset^3$ vectors in $\calG_{\SO}$ can be rearranged equivalently as a vector in $\Rset^6$, and vice versa. Moreover, the non-parallel and non-zero conditions defining $\calG_{\SO}$ are, in practice, numerically almost guaranteed to hold in finite-precision calculations. Overall, this allows $\calT_\theta^{-1}$ to generate unconstrained samples in a six-dimensional Euclidean space, while the rotation manifold constraints are subsequently enforced by the GSO transformation in $f_{\mathrm{GSO}}$.

With the change of variables in probability densities under the action of $\calT_\theta$, the conditional probability density of a rotation $\bfg$ (in GSO representation) can be written as
\be
\label{eq:cNF_density}
    p_{\theta}(\bfg|\bfc) 
    =
    p_z\!\left(\calT_{\theta}(\bfg;\bfc)\right)
    \left|\det \partderiv{\calT_{\theta}(\bfg;\bfc)}{\bfg} \right|.
\ee
Here, $\partial/\partial \bfg$ indicates the Jacobian with respect to the GSO coordinates of rotation, and $p_z$ is known analytically from \eqref{eq:latent_beta}. 

\textbf{Data:} Let $\calD=\{(\bfc_i,\bfR_i):\bfc_i\in\Rset^k; \bfR\in\SO, i=1,\dots,N\}$ denote a training dataset comprising condition-rotation pairs. The data may be obtained, for example, from EBSD measurements or CPFEM simulations and rotation matrices are projected onto the crystal's fundamental zone. For a single condition, such experiments or simulations may produce thousands to millions of rotation matrices, one for each grain. We simply associate the same condition with each rotation to construct the corresponding condition-rotation pairs. 

\textbf{Training:} The flow model is then trained by maximizing the likelihood (computed via \eqref{eq:cNF_density}) of the samples in the training dataset:
\be
\theta \leftarrow \argmaxWithArgs_\theta \prod_{i=1}^N p_\theta (g_\text{GSO}(\bfR_i) | \bfc_i),
\ee
which is equivalent to minimization of the negative log-likelihood:
\be
\theta \leftarrow \argminWithArgs_\theta \frac{-1}{N}\sum_{i=1}^N \log p_\theta (g_\text{GSO}(\bfR_i) | \bfc_i).
\ee

\textbf{Sampling:} For generative tasks, the objective is to generate a rotation for a given condition $\bfc$ such that it follows the conditional distribution $p_\theta(\bfg |\bfc)$. This is achieved by first sampling from the base distribution and then transforming the sample using $\calT_\theta^{-1}$:
\be
\bfR=f_\text{GSO}(\calT^{-1}_\theta(\bfz;\bfc))\qquad 
    \text{with}\quad
    \bfz\sim p_z(\bfz).
\ee

\textbf{Model architecture:}  We now discuss  the architecture of the invertible map $\calT_\theta$. The flow model is constructed as a composition of $L$ invertible bijectors:
\be
\label{eq:flow_comp}
    \calT_{\theta} = \left (T_{\theta_L}\circ T_{\theta_{L-1}}\circ\cdots\circ T_{\theta_1} \right)
    (\bfg;\bfc)
    \qquad \text{with}\quad \theta = \theta_L \cup \theta_{L-1} \cup\cdots\cup \theta_1,
\ee
where $T_{\theta_l}(\cdot;\bfc): \Rset^6 \rightarrow \Rset^6$ is the $l^\text{th}$ invertible bijective layer of the flow and $\theta_l$ is the corresponding set of trainable parameters. In the conditional setting, the condition $\bfc$ goes through each layer but is not itself transformed by the flow. Each layer of the flow is implemented using a neural spline coupling transform, specifically a rational–quadratic rolling spline coupling (RQ-RSC) transform. The RQ-RSC transform was first introduced by \citet{Durkan2019NeuralFlows}; here, we adopt the implementation of \citet{Crenshaw2024ProbabilisticFlows,Crenshaw2024Pzflow:V3.1.3}. We briefly review the RQ-RSC transform and refer the reader to these works for details of the implementation. 

For any coupling layer $T_{\theta_l}(\cdot;\bfc)$, let $\bfx\in\Rset^D$ and $\bfy\in\Rset^D$ denote the input and output, respectively. To ensure bijectivity, $\bfx$ and $\bfy$ must have the same dimensionality. Furthermore, because the input to the first layer and the output of the last layer are both six-dimensional, it follows that $D=6$.

The coupling layer splits the input into two blocks. The first $d$-dimensional block (with $d=3$ in our implementation) is passed through without any transformation, while the remaining $(D-d)$-dimensional block is transformed using a spline: 
\be
\label{eq:split_data}
\bfy = T_{\theta_l}(\bfx;\bfc) \  \left\{\quad 
\begin{aligned}
    \text{Pass-through:} \quad 
    & \bfy_{1:d} 
    = \bfx_{1:d},\\
    \text{Spline transform:} \quad 
    & \bfy_{d+1:D} 
    = \psi(\bfx_{d+1:D} | m_{ \theta_l} (\bfx_{1:d},\bfc)).
\end{aligned}
\right.
\ee
Here, $\psi(\bfx_{d+1:D}|\cdot)$ denotes a monotone rational-quadratic spline applied element-wise on coordinates $\bfx_{d+1:D}$.  The parameters of the spline $\psi$, including bin widths, bin heights, and derivatives, are obtained from a dense conditioner neural network $m_{\theta_l}$ with trainable parameters $\theta_l$. The conditioner neural network takes as inputs the pass-through block $\bfx_{1:d}$ and conditions $\bfc$. We use multi-layer perceptrons as the architecture for the conditioner neural network.

This coupling strategy ensures the invertibility of the transformation. Since the pass-through output remains unchanged, it can be combined with the known conditions to provide sufficient information to exactly reproduce spline parameters via the conditioner neural network. Given that the spline is monotonic, the entire transformation is therefore bijective, and the input can be recovered from the output as
\be
\label{eq:coupling_inverse}
\bfx = T^{-1}_{\theta_l}(\bfy;\bfc) \  \left\{\quad 
\begin{aligned}
    \text{Pass-through:} \quad 
    & \bfx_{1:d} = \bfy_{1:d},\\
    \text{Spline inverse:} \quad 
    & \bfx_{d+1:D} =
    \psi^{-1}
    \left(
        \bfy_{d+1:D} |
        m_{\theta_\ell}(\bfy_{1:d},\bfc)
    \right).
\end{aligned}
\right.
\ee
The invertibility of the coupling layer is ensured by construction, even though the conditioner neural network can be arbitrarily complex and need not be invertible.

An additional advantage of this coupling structure is that the Jacobian of the transformation, required in \eqref{eq:cNF_density}, becomes triangular. From \eqref{eq:split_data}, we obtain the Jacobian
\be
\label{eq:coupling_jacobian}
\begin{aligned}
    \frac{\partial \bfy}{\partial \bfx}
    =
    \begin{bmatrix}
        \bfI_{d} & \bfzero \\
        \dfrac{\partial \bfy_{d+1:D}}{\partial \bfx_{1:d}}
        &
        \dfrac{\partial \bfy_{d+1:D}}{\partial \bfx_{d+1:D}}
    \end{bmatrix},
\end{aligned}
\ee
where $\bfI_d$ is the $d \times d$ identity matrix. Consequently, the determinant of the Jacobian in the likelihood calculation \eqref{eq:cNF_density} only depends on the lower right block
\be
\label{eq:coupling_det}
\begin{aligned}
    \det
    \left(
    \frac{\partial \bfy}{\partial \bfx}
    \right)
    =
    \det
    \left(
    \frac{\partial \bfy_{d+1:D}}{\partial \bfx_{d+1:D}}
    \right),
\end{aligned}
\ee
which is straightforward to compute using \eqref{eq:split_data} and automatic differentiation.

\begin{figure}[htbp]
    \centering
    \includegraphics[width=\textwidth]{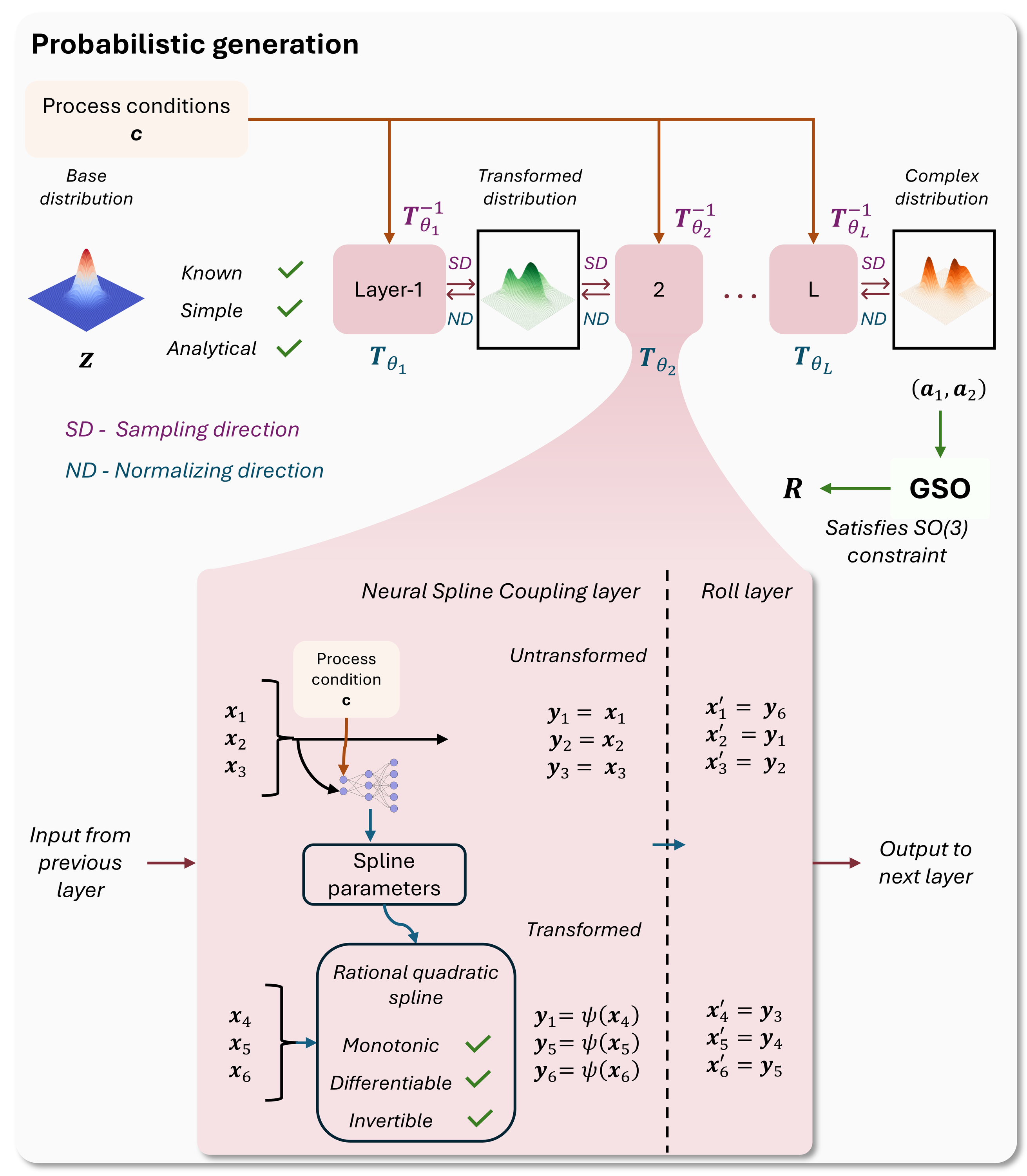}
    \caption{ 
    \textbf{Schematic illustration of TexFlow architecture}: for conditional probabilistic generation of rotations. Samples from a simple base distribution $\bfz$ are transformed through a sequence of flow layers, with the process conditions $\bfc$ provided to each layer. Each bijector comprises a rational quadratic neural spline coupling and a roll operation. The lower panel details one coupling layer, showing the use of untransformed components and conditions to predict the spline parameters followed by an invertible spline transformation of the remaining components. The final unconstrained vectors $(\bfa_1,\bfa_2)$ are converted into a valid rotation matrix $\bfR \in \SO $ through the GSO map.
    }
    \label{fig:nf_arch}
\end{figure}

A single coupling layer transforms only part of the input vector. To allow all of the $g_\text{GSO}$ coordinates to be transformed across the full model, the neural spline coupling layers are interleaved with rolling operations, which are cyclical permutation of the coordinates. In essence, the dimensions selected for pass-through and those subjected to spline transformation are permuted across layers.  Since a roll is a permutation, it is bijective with unit Jacobian determinant and doesn't contribute to the log-Jacobian determinant. The stacking of these coupling and rolling operations gives this architecture more expressivity. The latent distribution, presented in \eqref{eq:latent_beta}, is chosen to have support compatible with the spline range, while the conditional flow is obtained by supplying the condition $\bfc$ to the conditioner network in every coupling layer. In this way, $\bfc$ influences the learned spline parameters and therefore the conditional density $p_\theta(\bfg|\bfc)$.  Additional mathematical details of the neural spline flow are provided in ref. \cite{Crenshaw2024ProbabilisticFlows, Durkan2019NeuralFlows}.

\autoref{fig:nf_arch} provides an overview of the TexFlow model architecture. In summary, the model transforms samples from a simple base distribution $\bfz$ into a complex distribution of unconstrained GSO vectors $(\bfa_1, \bfa_2)$ through a series of conditional invertible flow bijector layers. The process conditions $\bfc$ are provided to the neural networks within each bijector and are used throughout the transformation. Each flow layer consists of a rational-quadratic neural spline coupling transformation followed by a roll operation. The lower part of the figure visualizes the operation of one such coupling layer. The input is divided into transformed and untransformed components, where the untransformed component and the conditions are used to predict the spline parameters for the transformed components. The transformed components are mapped using a monotonic, differentiable, and invertible rational quadratic spline. After passing through all flow layers, the resulting GSO vectors are mapped to valid rotation matrices using the Gram-Schmidt orthogonalization. This procedure enables conditional generation of rotations while preserving the constraints of the rotation manifold $\SO$, by construction.

\subsection{Hopf fibration}
\label{subsec:hopf_fibration}

Since texture is fundamentally a probability distribution over rotations on the non-Euclidean manifold $\SO$, it cannot be directly visualized in Euclidean space. One way to circumvent this limitation is to project the rotations to some other space where conventional visualization techniques can be applied. Ideally, such a projection should preserve as much information as possible about the underlying distribution on $\SO$. 

Pole figures and inverse pole figures are such techniques commonly used for crystallographic texture analysis. Typically, for a chosen pole direction $\bfu \in \Sii$, the rotation $\bfR\in\SO$ is projected to rotated pole $\bfR\bfu$. This projection is many-to-one, as multiple rotations can produce the same rotated pole, and therefore results in a loss of information. The resulting distribution of $\bfR\bfu$ on $\Sii$ is then visualized through a bijective stereographic projection $\Sii \rightarrow \Rset^2$, yielding the classical pole figure. The dimensionality of the spaces also makes this loss of information apparent: $\Sii$ or $\Rset^2$ used for visualization is two-dimensional, whereas the rotation space $\SO$ is three-dimensional.

\begin{figure}
    \centering
    \includegraphics[width=0.8\textwidth]{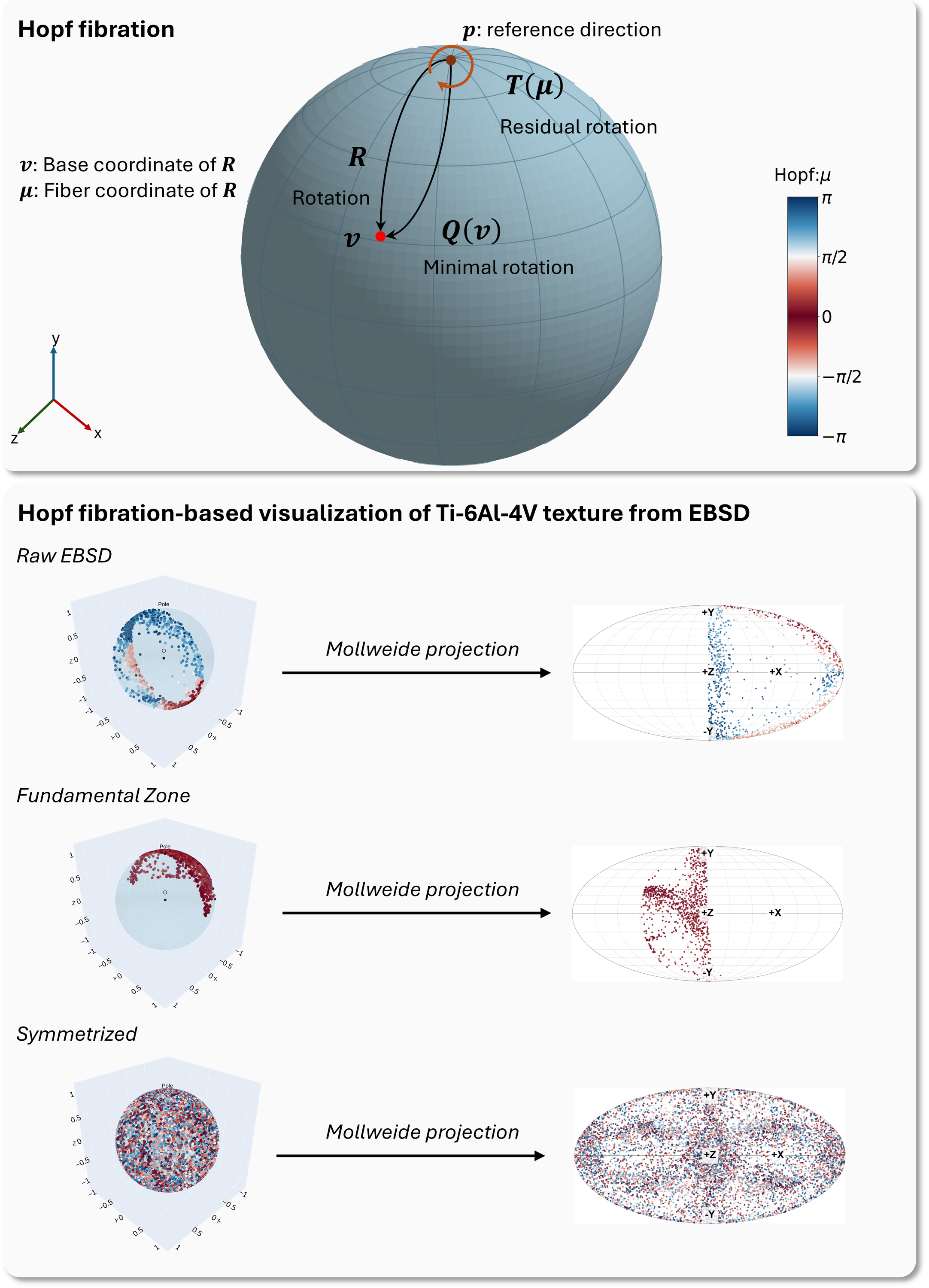}
    \caption{ \textbf{Schematic of Hopf fibration}. Top: A reference direction $\bfp\in \Sii$ is mapped by a rotation $\bfR$ to the base coordinate $\bfv = \bfR \bfp \in \Sii$. The minimal rotation $\bfQ(\bfv)$ maps $\bfp$ to $\bfv$, while the residual rotation $\bfT(\mu)$ acts about $\bfp$ as an axis and defines the fiber coordinate $\mu$. Thus, each rotation is represented by the pair $(\bfv, \mu)$, where $\bfv$ is represented spatially on $\Sii$, while the fiber coordinate $\mu$ is encoded by color.
    Bottom: Hopf maps are illustrated for Ti-6Al-4V EBSD data in the raw, crystallographic fundamental zone, and symmetrized forms. The three-dimensional Hopf map is accompanied by its Mollweide projection.}
    \label{fig:hopf_plot}
\end{figure}

Motivated by this, we propose a complementary technique for texture visualization based on the \emph{Hopf fibration} \citep{Hopf1931}. The Hopf fibration has been a landmark work in topology for many decades \cite{Lyons2022AnFibration}, but remains largely unexplored in materials science.
Here, we introduce a practical, materials-science-friendly reinterpretation of the Hopf fibration and refer readers interested in its topological foundations to refs.~\cite{Lyons2022AnFibration,Treisman2009AFibration}.  
\autoref{fig:hopf_plot} provides a schematic overview of the Hopf fibration. Before discussing the Hopf fibration, we first introduce some intermediate concepts. 

\textbf{Hopf map}: Let $\bfp\in S^2$ denote a reference direction (similar to a pole in pole figures). We choose the canonical $z$-axis as the reference direction, i.e., $\bfp=[0,0,1]^\intercal$.
For a given rotation $\bfR\in\SO$, the \textit{Hopf map} maps $\bfR$ to 
\be
\bfv = \bfR\bfp \in \Sii.
\ee
The rotations from $\bfp$ to $\bfv$ are not unique and infinitely many. The set of all such rotations constitutes a \textbf{Hopf fiber}: $\calF(\bfv)=\{\bfR\in\SO : \bfv=\bfR\bfp\}$, which has its own topological structure that is beyond the scope of this work. Since $\bfv$ identifies the fiber to which $\bfR$ belongs, we deem $\bfv$ as the base coordinate of $\bfR$.

\textbf{Minimal rotation:} Among all the rotations in a fiber $\calF(\bfv)$ associated with $\bfv$, let $\bfQ(\bfv)\in\calF(\bfv)$ denote the \emph{minimal rotation}, defined as the rotation about the axis $\bfk$ normal to the plane spanned by $\bfp$ and $\bfv$, with the rotation angle $\theta$ equal to the angle between $\bfp$ and $\bfv$. Knowing the rotation axis and rotation angle by definition, $\bfQ$ is given by
\be
\bfQ(\bfv) = \bfI + (\sin \theta)[\bfk]_\times + (1-\cos\theta)[\bfk]^2_\times, \quad \text{with}\quad \bfk = \bfp\times\bfv \quad\text{and}\quad \theta = \cos^{-1}(\bfp\cdot\bfv),
\ee
where we used Rodrigues' formula \eqref{eq:axis_angle_short} for the first term. Since both $\bfR$ and $\bfQ(\bfv)$ belong to $\bfF(\bfv)$, they satisfy the relation:
\be\label{eq:hopf_map_equivalenceQ}
\bfv = \bfR\bfp = \bfQ(\bfv)\bfp.
\ee

\textbf{Residual rotation:} We decompose $\bfR$ into a composition of two rotations:
\be\label{eq:hopf_decomp}
\bfR = \underbrace{\bfQ(\bfv)}_{\substack{\text{minimal}\\\text{rotation}}} \ \underbrace{\bfT(\mu)}_{\substack{\text{residual}\\\text{rotation}}},
\ee
where $\bfT(\mu)\in\SO$ denotes a \textit{residual rotation} with rotation angle $\mu\in(-\pi,\pi]$.  Substituting the decomposition into \eqref{eq:hopf_map_equivalenceQ} results in
\be 
\bfQ(\bfv)\bfT(\mu)\bfp = \bfQ(\bfv)\bfp.
\ee

Multiplying $\bfQ^\intercal(\bfv)$ on both sides yields a constraint on the residual rotation:
\be
\bfT(\mu)\bfp = \bfp.
\ee

This implies that $\bfp$ is invariant under the residual rotation $\bfT(\mu)$, which is only possible if $\bfp$ is itself the axis of rotation of $\bfT(\mu)$.
Therefore, $\bfT(\mu)$ is given by (using \eqref{eq:axis_angle_short}):
\be
\bfT(\mu) = \bfI + (\sin \mu)[\bfp]_\times + (1-\cos\mu)[\bfp]^2_\times,
\ee

where $\mu$ is the angle of rotation. 
Since $\bfp=[0,0,1]^\intercal$ in our case, $\bfT(\mu)$ further simplifies as
\be
\label{eq:residualT_simplified}
\bfT(\mu)=\begin{pmatrix}
    \cos \mu & -\sin\mu & 0 \\
    \sin \mu & \cos \mu & 0 \\
    0 & 0 & 1
\end{pmatrix}.
\ee
From \eqref{eq:hopf_decomp} and \eqref{eq:residualT_simplified}, $\mu$ can be calculated as
\be
\mu = \text{atan2}(T_{22}, T_{11})\quad \text{with} \quad \bfT(\mu) = \bfR\bfQ^\intercal(\bfv),
\ee
where $\text{atan2}$ denotes the 2-argument arctangent. The parameter $\mu$ uniquely distinguishes the rotations within the fiber $\calF(\bfv)$ and is therefore deemed as the fiber coordinate.

In summary, any rotation $\bfR$ can be represented by two Hopf coordinates:
\be
\begin{aligned}
    \text{\textit{Base coordinate} in $\Sii$:} & \quad \bfv = \bfR\bfp\\
    \text{\textit{Fiber coordinate}  in $\Si$:} & \quad \mu = \text{atan2}(T_{22}, T_{11})\quad \text{with} \quad \bfT(\mu) = \bfR\bfQ^\intercal(\bfv).
\end{aligned}
\ee
In the Hopf plots, the base coordinate $\bfv$ is plotted as a point on the unit sphere (or its Mollweide projection), while the fiber coordinate $\mu$ is encoded as the color. As an example, \autoref{fig:hopf_plot} visualizes the crystallographic texture of Ti-6Al-4V obtained from EBSD using Hopf fibration.

Let $\mathcal{T}_{\mathrm{cr}}\subset\SO$ denote the crystallographic point symmetry group (e.g., cubic $m\bar{3}m$ or hexagonal $6/mmm$). Two rotations $\bfR$ and $\bfR\bfG$ for $\bfG\in\mathcal{T}_{\mathrm{cr}}$  are equivalent due to crystal symmetries. Therefore, the physically distinct rotation space is the quotient $\SO/\mathcal{T}_{\mathrm{cr}}$ \cite{Man2022CrystallographicRepresentations}. All the Hopf plots in this work visualize the rotation distribution on the quotient space $\SO/\calT_{\text{cr}}$, i.e., each rotation is visualized \emph{after projection into the crystallographic fundamental zone} using the crystal's symmetry group $\calT_{\text{cr}}$ to eliminate redundant copies of equivalent rotations. \autoref{fig:hopf_plot} illustrates this procedure for the Ti-6Al-4V EBSD data by comparing the raw texture data, the data reduced to the crystallographic fundamental zone, and the data expanded over all symmetry-equivalent rotations (referred to as ``symmetrized'' in the figure).

\begin{figure}
    \centering
    \includegraphics[width=\textwidth]{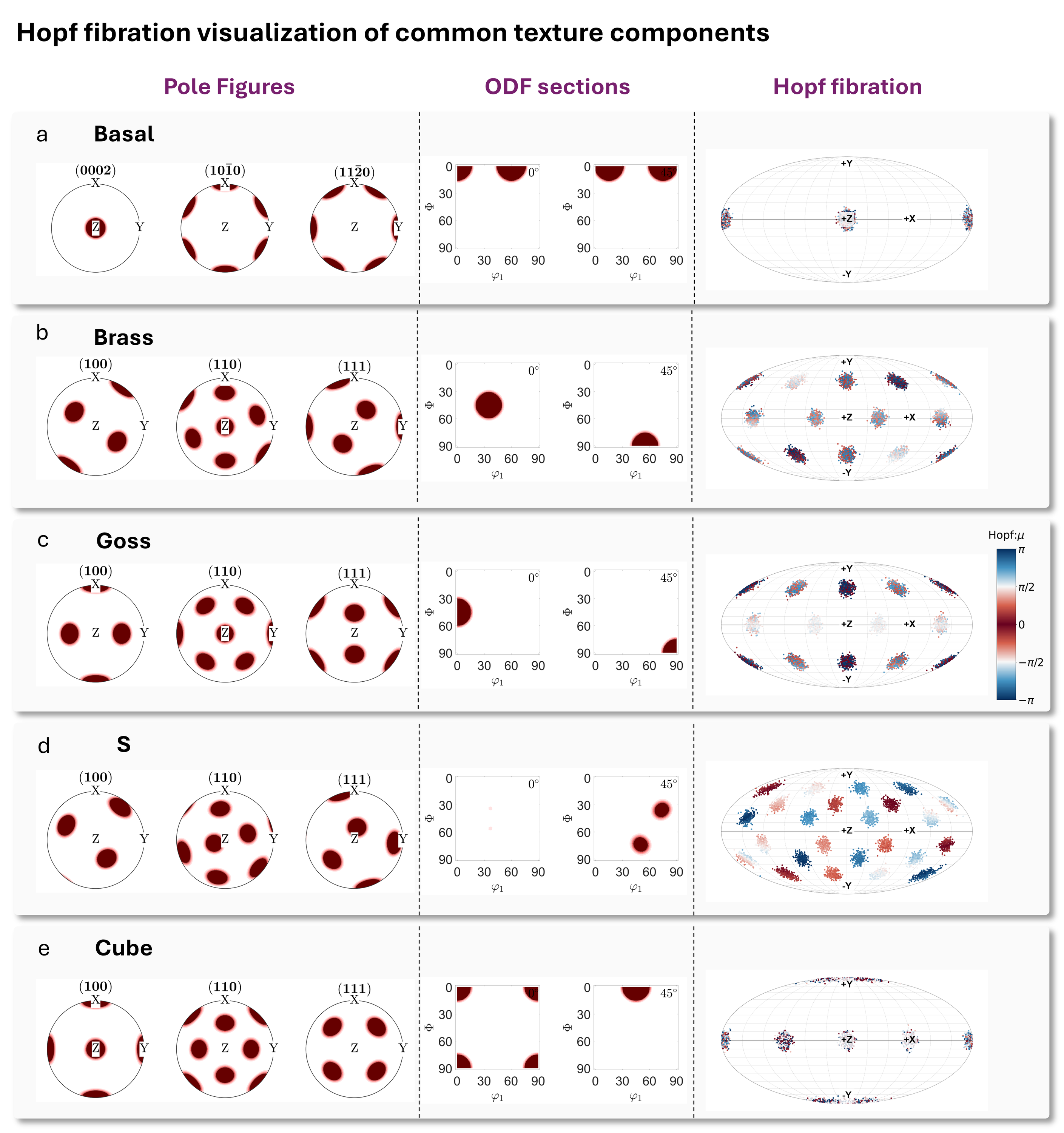}
    \caption{
    \textbf{Hopf fibration plot of representative crystallographic textures.} For each texture component, conventional pole figures are 
    shown in the left column, ODF sections at $\varphi_2 =0^\circ$ and $45^\circ$ are shown in the center, and the Hopf maps are shown in the right column. The texture examples include (a) basal $\{0001\}$ texture, (b) Brass $\{110\}\langle112\rangle$, (c) Goss $\{110\}\langle001\rangle$, (d) S $\{123\}\langle634\rangle$ and (e) Cube $\{001\}\langle100\rangle$. }
    
    \label{fig:hopf_common}
\end{figure}

\textbf{Hopf fibration of common textures:} In this work, the Hopf fibration is used as a complementary visualization tool alongside conventional pole figures and ODF sections. \autoref{fig:hopf_common} illustrates the representation of several canonical crystallographic textures, alongside their conventional pole figures and ODF sections, to build intuition for interpreting Hopf plots.

A Hopf plot is similar to a pole figure in the sense that the base coordinate is obtained from the projection of a pole direction. Unlike a pole figure, however, the Hopf representation retains an additional dimension through the fiber coordinate. Thus, both the rotation manifold and the Hopf coordinates are three-dimensional, allowing the Hopf representation to retain the full information of the rotation. At the same time, the Hopf plot retains geometric features of the texture that are readily interpretable (similar to pole figures) but are not immediately apparent in Euler-angle ODF sections. It is particularly valuable for comparing ground truth textures with samples generated by proposed TexFlow model, as it provides a global view of  multimodal and continuous texture  directly on $\SO$ in a single plot.

\begin{figure}[t]
    \centering
    \includegraphics[width=\textwidth]{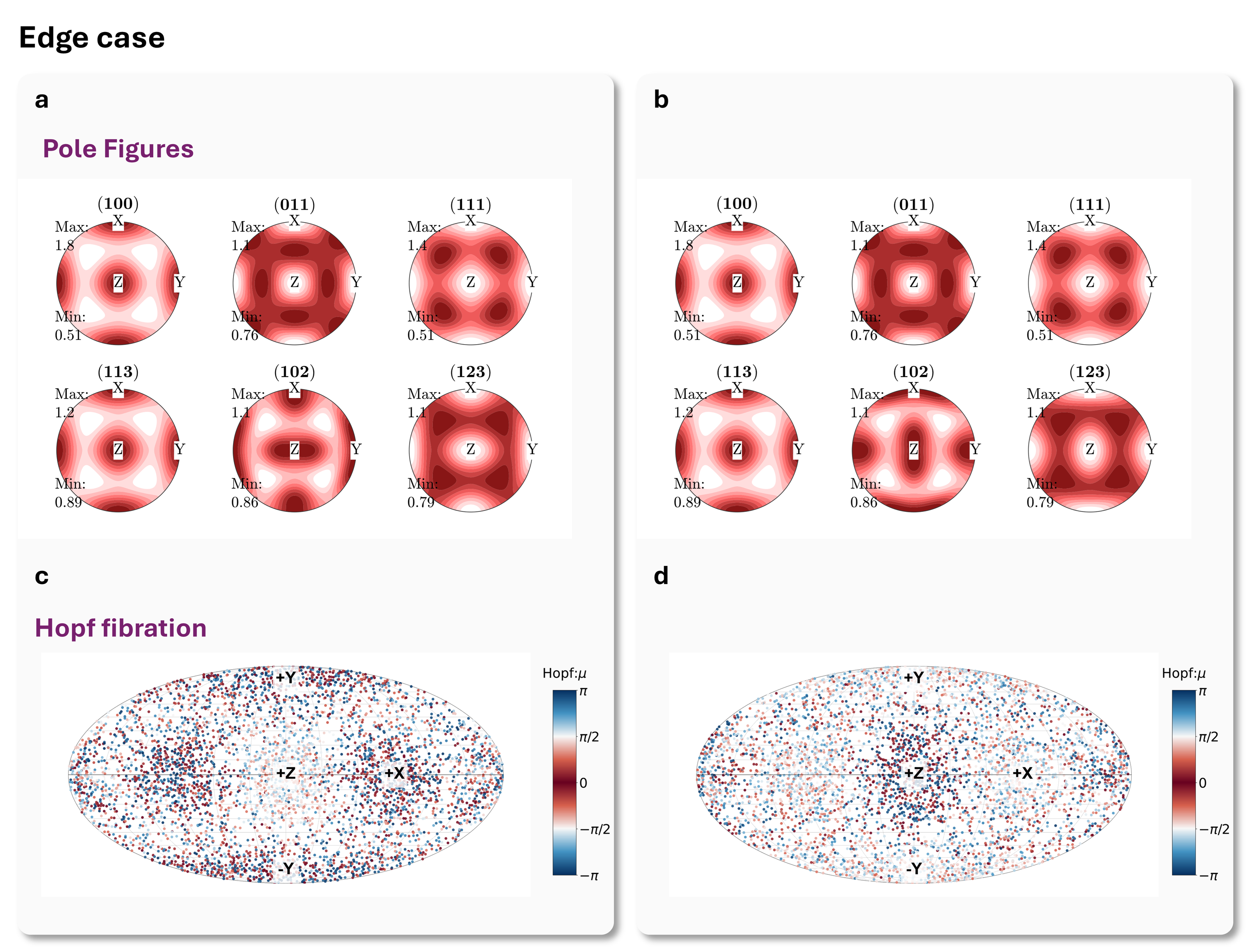}
    \caption{
    Illustration of the ambiguity that can arise in pole figures representation of textures. Panels (a) and (b) show the pole figures of two distinct textures. The pole figures corresponding to $ (100), (011), (111), (113)$ are indistinguishable, whereas differences are only observed for poles $(102)$ and $(123)$. Panels (c) and (d) show the corresponding Hopf maps of the two textures. Unlike the conventional pole figures, the Hopf maps produce distinct patterns for the two textures.
    }
    \label{fig:pf_ambiguity}
\end{figure}

To further highlight the capabilities of the Hopf fibration relative to pole figures, \autoref{fig:pf_ambiguity} illustrates two distinct textures whose pole figures for nearly all of the considered poles, including $(100)$, $(011)$, $(111)$, and $(113)$, are identical. The ambiguity can only be resolved by considering specific pole figures, such as $(102)$ and $(123)$. This example was first proposed by \citet{Matthies1979OnPhenomena} and further discussed by refs.~\cite{hielscher_mtex_pf, Schaeben2025CommentRepresentation}. In contrast to pole figures, the Hopf representation produces markedly different patterns for the two textures in a single plot, immediately resolving the ambiguity because no information is lost in the visualization.

\section{Results}
\label{sec:results}

We evaluate the TexFlow framework for generative tasks on three types of data: (i) unconditional data obtained from EBSD, (ii) conditional data obtained from EBSD, and (iii) conditional data obtained from CPFEM simulations. \autoref{tab:dataset_summary} summarizes the datasets used, with additional details provided in the discussion of the individual tasks. For the unconditional tasks, the objective is to learn and sample the probability distribution of the rotation data, i.e., the texture. For the conditional tasks, the objective is to learn and sample the conditional probability distribution of the rotation data given the process parameters and to predict the corresponding texture for previously unseen process parameters.

\begin{table}[t]
\centering
\caption{Summary of datasets used for training and evaluation of the TexFlow framework.}
\label{tab:dataset_summary}
\footnotesize
\renewcommand{\arraystretch}{1.4} %
\begin{tabular}{@{} 
    >{\raggedright\arraybackslash}m{2.0cm} 
    >{\raggedright\arraybackslash}m{2.0cm} 
    >{\raggedright\arraybackslash}m{2.4cm} %
    >{\raggedright\arraybackslash}m{3.0cm} 
    >{\raggedright\arraybackslash}m{2.9cm} 
    c @{}}
\toprule
\textbf{Dataset type} &
\textbf{Material} &
\textbf{Symmetry} &
\textbf{Conditions} &
\textbf{Data size}$^{\mathrm{a}}$ &
\textbf{Source ref.} \\
\midrule

\multirow{3}{*}{\makecell[l]{Unconditional\\EBSD}} 
& AM SS316L & Cubic ($m\bar{3}m$) & -- & 2,305 
& \cite{Agrawal2020High-throughputSteel} \\
\cmidrule(l){2-6}

& Forsterite & Orthorhombic ($mmm$) & -- & 1,032 
& \cite{Bachmann2010TextureToolbox} \\
\cmidrule(l){2-6}

& Glaucophane & Monoclinic ($2/m$) & -- & 8,840 
& \cite{Ott2024SeismicRecord} \\
\midrule
 
\makecell[l]{Conditional\\EBSD} 
& \makecell[l]{Hot-rolled\\Ti-6Al-4V} 
& Hcp ($6/mmm$)
& Temperature: 9 values, \newline \% reduction: 4 values, \newline directions: 2 (RD/TD)
& $100{,}000$  per condition; not all conditions available; \newline No. of conditions: 56
& \cite{Zeng2022EBSDTi-6Al-4V,Zeng2025EffectRolling} \\
\midrule

\makecell[l]{Conditional\\CPFEM} 
& Copper 
& Cubic ($m\bar{3}m$)
& Timesteps: 50 values
& $10{,}000$  per condition
& In-house \\  
\bottomrule
\end{tabular}

\begin{minipage}{\textwidth}
\vspace{2mm}
\footnotesize
$^{\mathrm{a}}$ Here, data size means the number of rotation matrices.\\
\end{minipage}
\end{table}

For clarity, results from the \textbf{ground truth} and the \textbf{TexFlow} model are denoted by the abbreviations \textbf{GT} and \textbf{TF}, respectively, throughout the subsequent discussion. Across all cases, comparisons with ground-truth data are reported using conventional pole figures and ODF section plots, complemented by Hopf plots to assess the learned distribution. More quantitative comparisons between the GT and TF distributions are performed using three quantities:
\begin{itemize}
\item \textit{Mean Marginal Kullback-Leibler divergence (MM-KLD):} Let $\calP=\{\bfR_i\}_{i=1}^{N}$ and $\calQ=\{\bfR_i\}_{i=1}^{M}$ denote $N$ and $M$ samples drawn from the GT data and TF distribution, respectively. We compute the empirical discrete marginal  probability for each component of $\bfR$ using $B>0$ histogram bins $\{\gamma_b\}_{b=1}^{B+1}$:
\be
p_{b,kl}(\calP) = \frac{1}{N}\sum_{i=1}^N \mathbbm{1}\left[\gamma_b\leq R_{i,kl}<\gamma_{b+1}\right] \qquad \text{and} \qquad 
q_{b,kl}(\calQ) = \frac{1}{M}\sum_{i=1}^M \mathbbm{1}\left[\gamma_b\leq R_{i,kl}<\gamma_{b+1}\right],
\ee
for GT and TF, respectively. Here, $b$ denotes the $b^\text{th}$ bin, $k,l\in\{1,2,3\}$ denote the $k^\text{th}$ row and $l^\text{th}$ column of a $3\times3$ matrix, respectively, and $\mathbbm{1}[\cdot]$ denotes the indicator function which equals 1 when its argument is true and 0 otherwise. The Kullback-Leibler divergence measures how much an approximating distribution ($q_{b,kl}$ of TF) differs from the true distribution  ($p_{b,kl}$ of GT). We define the MM-KLD as the mean of the marginal Kullback-Leibler divergences over all components of $\bfR$, i.e., 
\be
\label{eq:kld}
\text{MM-KLD} = \frac{1}{9}\sum_{k=1}^3\sum_{l=1}^3  \sum_{b=1}^{B} p_{b,kl} \log\left( \frac{p_{b,kl}}{q_{b,kl}}\right).
\ee

\item \textit{Mean Marginal Earth Mover's Distance (MM-EMD):} Let $\tilde\calP_{kl}=\text{Sort}\left(\{R_{i,kl}\}_{i=1}^N\right)$ and $\tilde\calQ_{kl}=\text{Sort}\left(\{R_{i,kl}\}_{i=1}^M\right)$ denote the sorted sets of $k^\text{th}$ row $l^\text{th}$ column components of rotation samples from GT data and TF distribution, respectively. 
We define MM-EMD as the earth mover's distance (also known as the one-dimensional Wasserstein distance) between the GT and TF marginal distributions averaged across all the components of $\bfR$:
\be
\label{emd}
\text{MM-EMD}
=\frac{1}{9}\sum_{k=1}^3\sum_{l=1}^3 \frac{1}{\tilde N}\sum_{i=1}^{\tilde N}
\left|x_{i,kl} - y_{i,kl}\right|,
\qquad \text{with}\quad
x_i\in\tilde\calP_{kl}
,\quad
y_i\in\tilde\calQ_{kl},
\ee
and $\tilde N=\min(N,M)$.

\item \textit{Texture index:} (also called $J$-index) is a scalar measure of the overall strength of the texture. For a given ODF $f(\bfR)$, it is defined as
\be
\label{eq:texture_index}
J(f) = \int_{\SO} f(\bfR)^2 \dd\mu(\bfR).
\ee
A uniformly random texture has $f(\bfR) = 1$ and $J=1$, whereas larger values indicate concentration of orientations and a ``sharper'' texture \cite{hielscher_mtex_J}. In this work, we compare the texture index of GT data and TF distribution. 
\end{itemize}

\textbf{Implementation details}: We adapt the Neural Spline Flow implementation from the \texttt{PZFlow} package \citep{Crenshaw2024Pzflow:V3.1.3} which leverages efficient GPU-accelerated training. The complete set of architectural choices and hyperparameters (e.g., learning rate, optimizer, batch size, etc.) is summarized in \autoref{tab:training_hyper}. As a preprocessing step for both learning and visualization, all rotations are reduced to a unique representation within the crystallographic fundamental zone using standard functions available in the MTEX toolbox \citep{Bachmann2010TextureToolbox}.

\subsection{Case 1: Unconditional generation on experimental EBSD data} 
\label{subsec:res_uncondNF}

We consider three texture datasets obtained via EBSD:
\begin{itemize}
\item additively manufactured (AM) stainless steel SS316L with cubic $m\bar{3}m$ symmetry and 2,305 data points, originally from \citet{Agrawal2020High-throughputSteel} and accessed via Ref.~\cite{Wanni2024MachineRepresentation};
\item forsterite with orthorhombic ($mmm$) symmetry and 1,032 data points, originally from \citet{Bachmann2010TextureToolbox} and accessed via Ref.~\cite{Wanni2024MachineRepresentation};
\item glaucophane with monoclinic ($2/m$) symmetry and 8,840 data points, originally from \citet{Ott2024SeismicRecord} and accessed via Ref.~\cite{Wanni2024MachineRepresentation}.
\end{itemize}

\begin{figure}
    \centering
    \includegraphics[width=\textwidth]{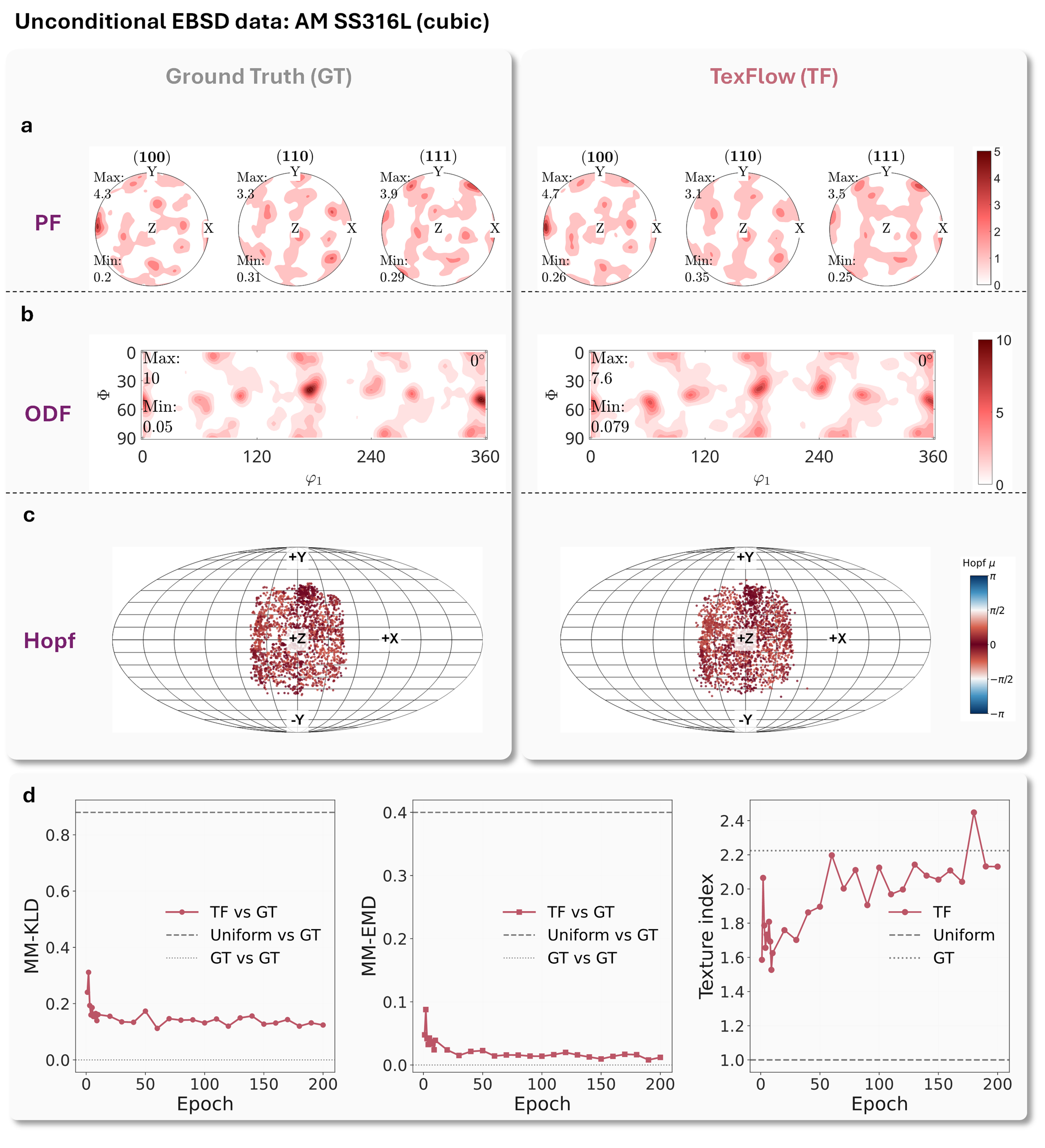}
    \caption{ 
     \textbf{Comparison of TexFlow vs.~ground truth on unconditional EBSD data of AM SS316L (cubic)}: (a) pole figures (PF), (b) ODF sections, (c)  Hopf plots, and (d) evolution of MM-KLD, MM-EMD, and texture index with training epochs.
    }
    \label{fig:uncond_ss316}
\end{figure}

\autoref{fig:uncond_ss316} presents the results of an unconditional TexFlow model trained on the AM SS316L dataset. The pole figures (\autoref{fig:uncond_ss316}a) indicate that the TF samples reproduce the GT cubic texture of AM SS316L, with intensity clusters near the ${100}$ and ${110}$ poles aligned with build direction. Texture intensity is interpreted as a multiple of random distribution (MRD) with  good agreement in  peak strengths between GT and TF samples. The texture features are consistent with solidification-controlled grain growth in laser-based AM, where preferential $\langle001\rangle$ and $\langle101\rangle$ fibers aligned with build direction are commonly observed \citep{Ishimoto2020CrystallographicMelting}. In addition to the dominant components, both GT and TF pole figures exhibit a diffuse background intensity, reflecting the broad orientation spread associated with remelting. The corresponding ODF sections (\autoref{fig:uncond_ss316}b) further show that TexFlow captures the broadened cubic texture peaks along with a continuous distribution, rather than producing artificially sharp or isolated components. This indicates the model preserves both dominant orientations and the experimentally observed dispersion in AM SS316L. In  \autoref{fig:uncond_ss316}c, we observe the global view of the texture in the Hopf fibration plots. In Hopf map, the TF samples form clusters that closely mimic those of the GT distribution. 

Complementing the qualitative comparisons, \autoref{fig:uncond_ss316}d shows the evolution of MM-KLD, MM-EMD, and texture index during the training. For reference, we report the MM-KLD and MM-EMD for a uniformly random texture relative to the GT data, providing a baseline representative of a model that fails to learn the underlying texture distribution. We also report the texture index of a uniformly random texture, which is equal to one, as a reference for the corresponding failed-model baseline. Additionally, we report the MM-KLD and MM-EMD of the GT data relative to itself, both of which are zero and represent the ideal case of a perfectly accurate model.

Both MM-KLD and MM-EMD decrease during the training and subsequently remain far below the corresponding uniform random reference values and close to zero. The texture index also evolves during training and gradually approaches the GT value as training proceeds. Overall, the qualitative and quantitative metrics indicate that TexFlow captures the features of the GT distribution, including its non-uniformity, peaks and overall texture strength.

\begin{figure}
    \centering
    \includegraphics[width=0.7\textwidth]{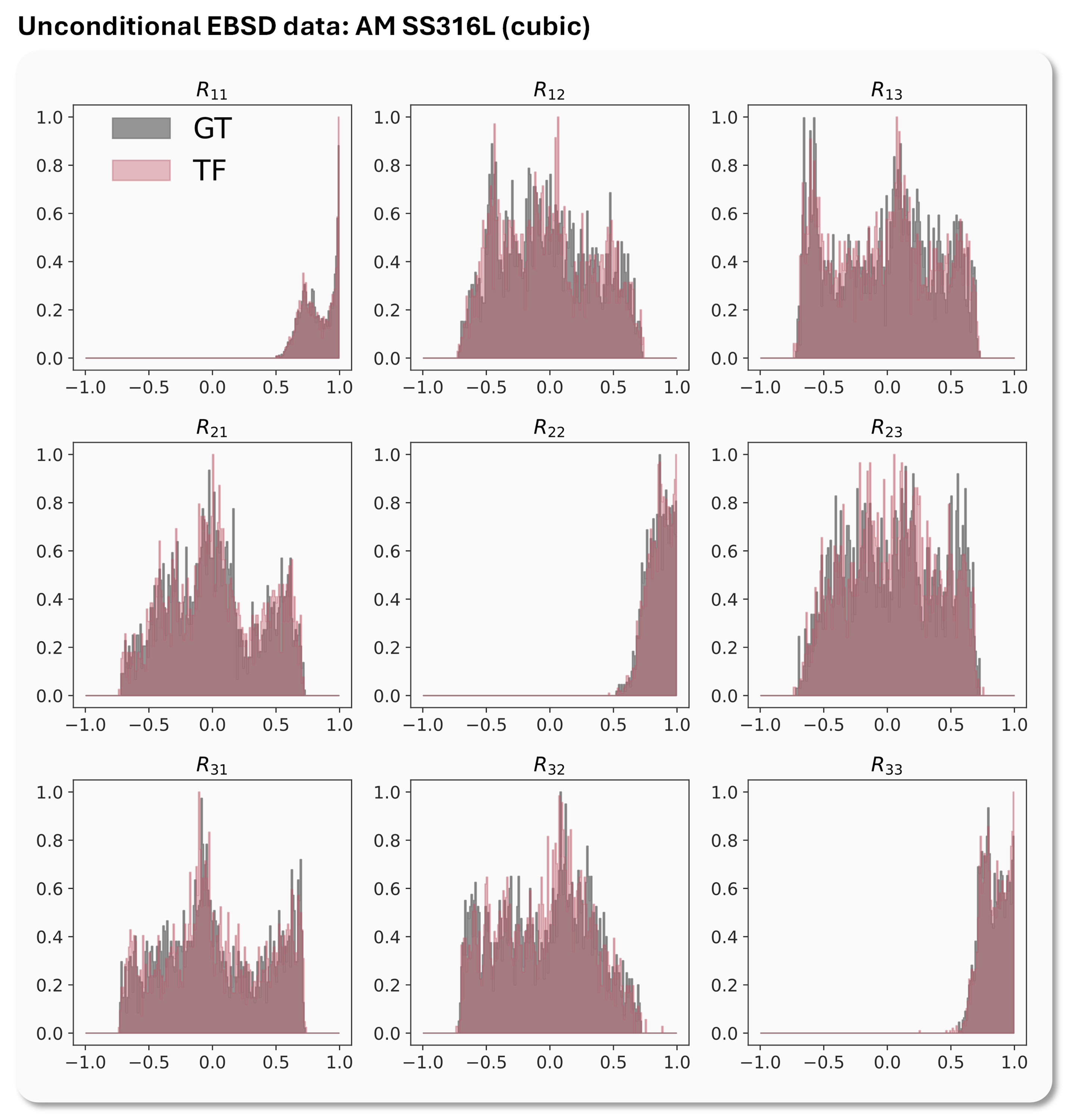}
    \caption{
    \textbf{Comparison of TexFlow vs.~ground truth on unconditional EBSD data of AM SS316L (cubic)}: marginal distributions of the nine components of the rotation matrices.
    }
    \label{fig:marg_ss316l}
\end{figure}

\begin{figure}
    \centering
    \includegraphics[width=0.85\textwidth]{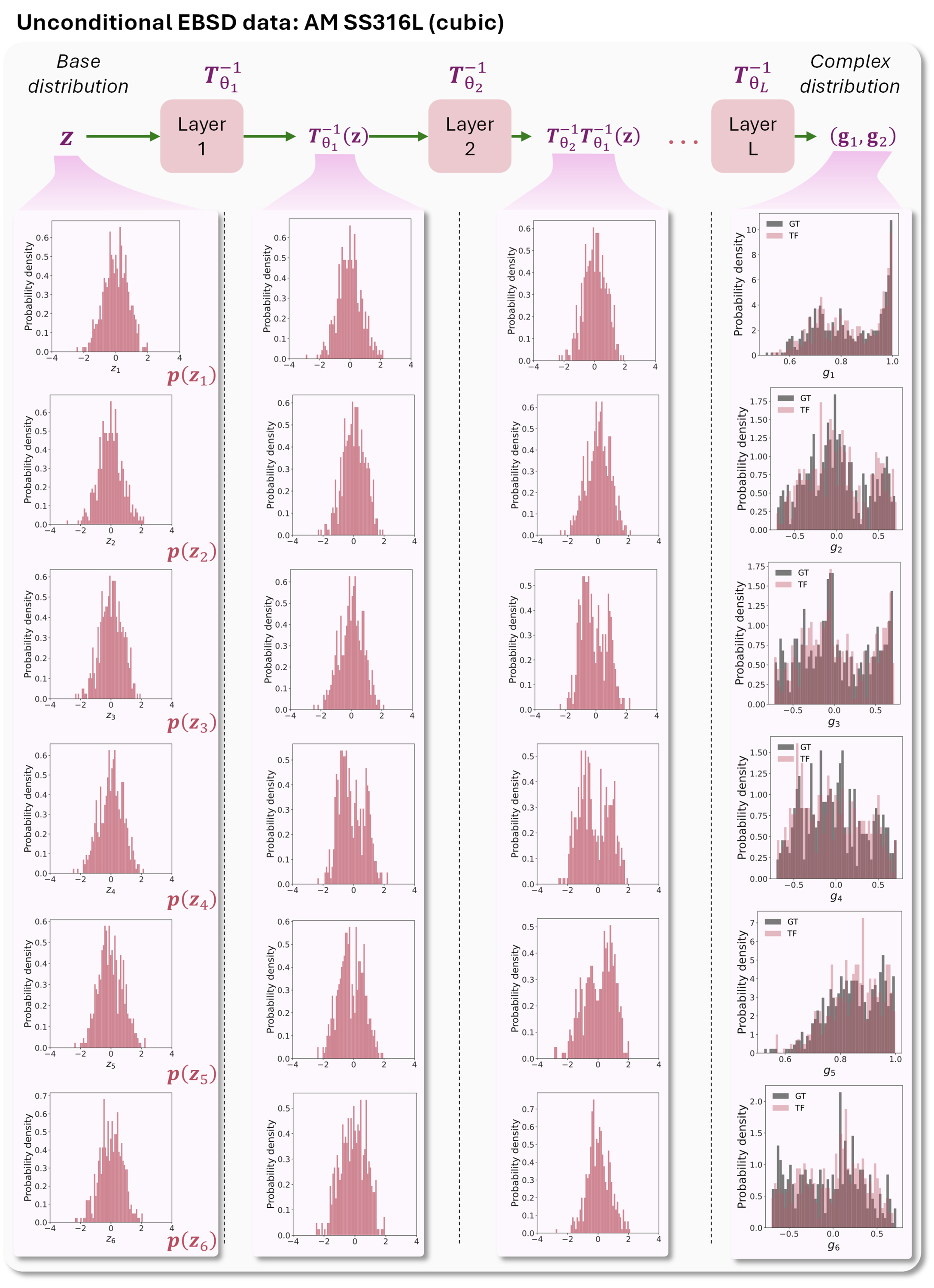}
    \caption{ 
     \textbf{Evolution of marginal distributions across TexFlow layers}: Samples from the base distribution of $\bfz$ are progressively transformed through the flow layers $T_{\theta_1}^{-1}$, $T_{\theta_2}^{-1}$, \dots, $T_{\theta_L}^{-1}$. Each column shows the marginal distributions after a particular layer, progressing from the simple base distribution in the leftmost column to the target distribution of the GSO coordinates in the rightmost column.
    }
    \label{fig:evolve_texflow}
\end{figure}

\begin{figure}
    \centering
    \includegraphics[width=\textwidth]{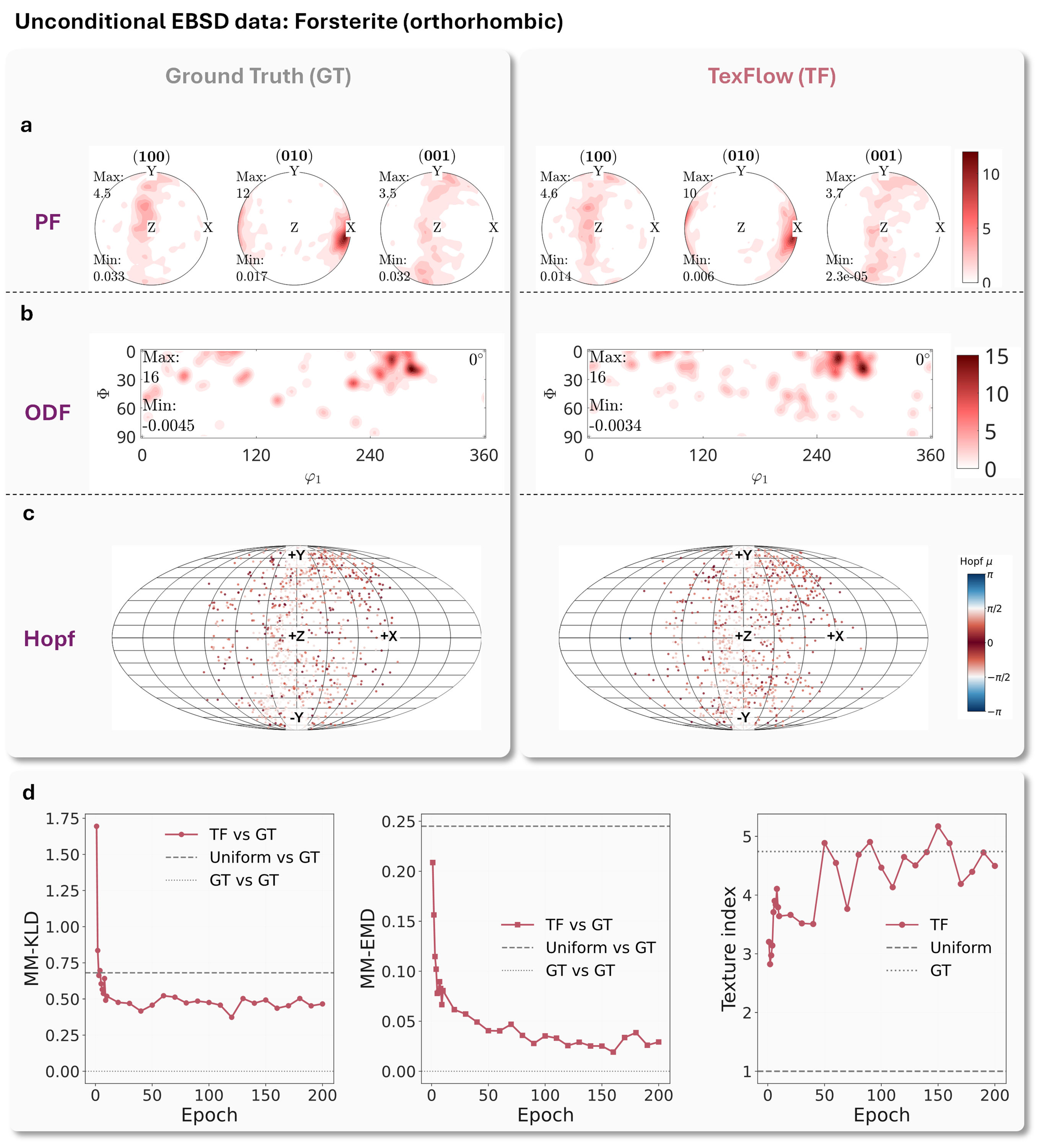}
    \caption{
    \textbf{Comparison of TexFlow vs.~ground truth on unconditional EBSD data of forsterite (orthorhombic)}: (a) pole figures (PF), (b) ODF sections, (c)  Hopf plots, and (d) evolution of MM-KLD, MM-EMD, and texture index with training epochs.
    }
    \label{fig:uncond_fors}
\end{figure}

\begin{figure}
    \centering
    \includegraphics[width=\textwidth]{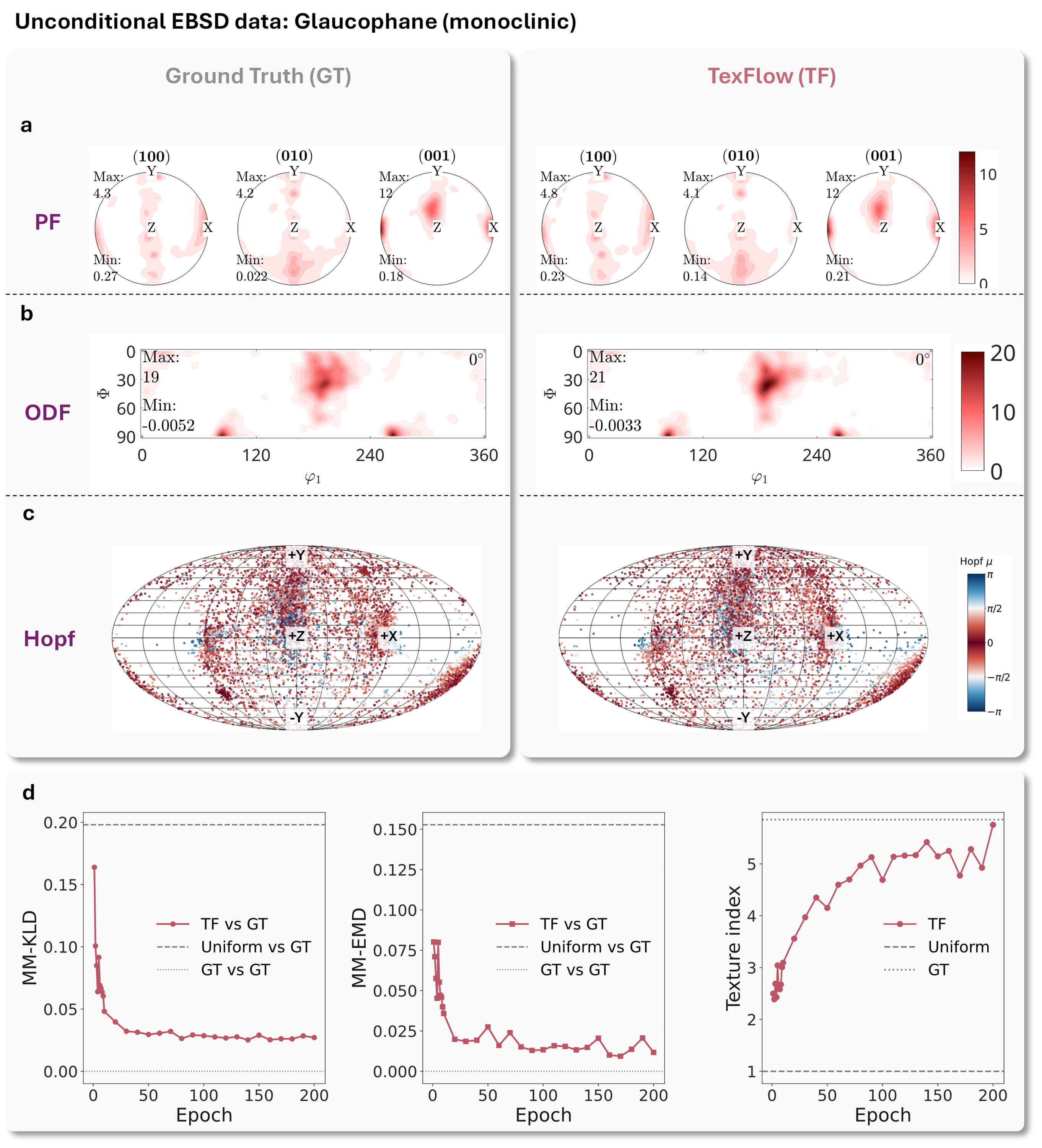}
    \caption{ 
    \textbf{Comparison of TexFlow vs.~ground truth on unconditional EBSD data of glaucophane (monoclinic)}: (a) pole figures (PF), (b) ODF sections, (c)  Hopf plots, and (d) evolution of MM-KLD, MM-EMD, and texture index with training epochs.
    }
    \label{fig:uncond_glau}
\end{figure}

As the final GSO samples are mapped back to rotation matrices, we further compare the marginal distributions of the nine matrix components in \autoref{fig:marg_ss316l}. The TF distributions reproduce the peaks and non-uniform spread observed in the GT distribution across $R_{11},\ldots,R_{33}$. The agreement is evident for components with strongly concentrated values near $\pm 1$, as well as for components exhibiting broader or multimodal distributions. The comparison demonstrates that TexFlow preserves the characteristic marginal distribution of the rotations except for small discrepancies in fine scale features.

To examine how the flow model transforms the distribution during training, we visualize the evolution of the orientation distribution through successive inverse transformations in \autoref{fig:evolve_texflow}. The first column (leftmost) shows the six components of the base latent variable $\bfz = (z_1, \ldots, z_6)$, which approximately follow the prescribed centered Beta distribution. The subsequent columns show the corresponding marginal distribution after applying the inverse of the first and second layers of the flow transformation, denoted by $T_{\theta_1}^{-1}$ and $T_{\theta_2}^{-1}$, respectively. Each coupling layer progressively modifies and warps the initially simple base distribution. The final stage, corresponding to the complete six-layer flow used for this illustration, produces the learned distribution in the GSO representation. The final (rightmost) column shows good agreement between the TF and corresponding GT marginal distributions of the GSO coordinates $\bfg = (g_1, \ldots, g_6)$. Together with the low MM-KLD and MM-EMD values, this confirms that the base distribution has been successfully transformed into the distribution of the observed data.

Not limiting ourselves to cubic symmetry or AM SS316L, we extend the same analysis to the orthorhombic Forsterite and monoclinic Glaucophane cases in \autoref{fig:uncond_fors} and \autoref{fig:uncond_glau}, respectively. In both cases, we observe good agreement between the TF and GT textures, both qualitatively through pole figures, ODF sections, and Hopf plots and quantitatively through MM-KLD, MM-EMD, and texture index. Together, these results demonstrate that the unconditional TexFlow model can learn multimodal experimentally obtained crystallographic textures across different crystal classes.

\subsection{Case 2: Conditional generation on experimental EBSD data} 
\label{subsec:res_condNF_ebsd}

\begin{figure}
    \centering
    \includegraphics[width=0.9\textwidth]{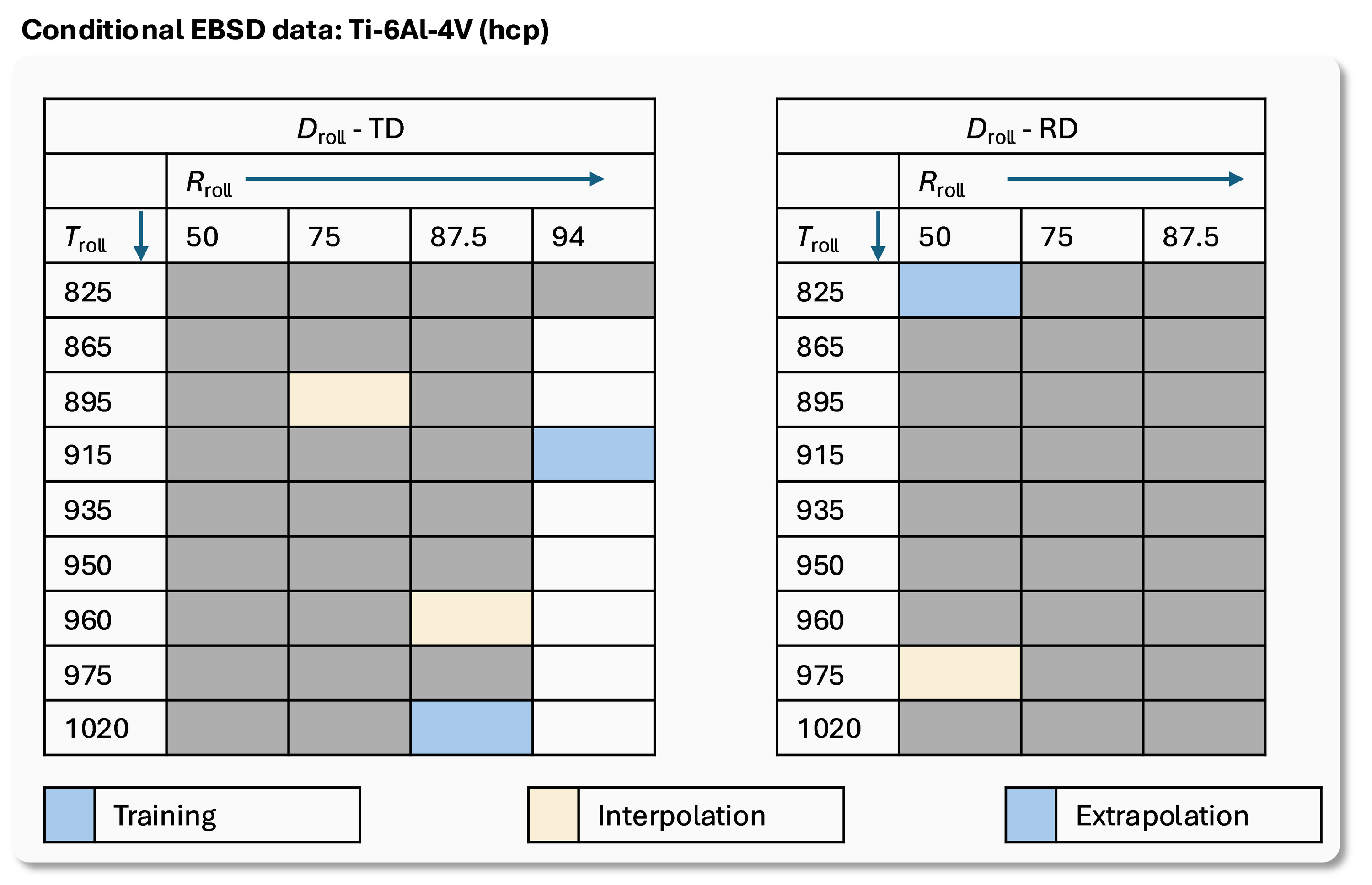}
    \caption{ 
    \textbf{Dataset split for conditional EBSD texture modeling}: The Ti-6Al-4V dataset is partitioned by rolling temperature ($\Temp$), percent rolling reduction ($\Roll$), and acquisition direction ($\Dir$). Grey cells denote training conditions; light yellow and light blue cells indicate validation conditions which are withheld to evaluate interpolation and extrapolation, respectively. 
    }
    \label{fig:data_split}
\end{figure}

We now consider a conditional generation task in which the TexFlow model is trained on an EBSD dataset of hot-rolled titanium alloy (Ti-6Al-4V) from refs.~\cite{Zeng2022EBSDTi-6Al-4V,Zeng2025EffectRolling}. This dataset shows the texture evolution of $\alpha$-phase (hcp) during hot rolling process, including development and variation of prismatic $\{11\bar{2}0\}\langle 10\bar{1}0\rangle$  components with  deformation and temperature. The conditioning variables are $\bfc=(\Roll,\Temp,\Dir)$, where $\Roll$ denotes the rolling reduction in percentage, $\Temp$ the rolling entry temperature, and $\Dir\in\{\RD,\TD\}$ the acquisition direction, with $\RD$ and $\TD$ corresponding to the rolling and transverse directions, respectively. The EBSD dataset spans four rolling reductions: $\Roll\in\{50\%,75\%,87.5\%,94\%\}$ and nine rolling entry temperatures: $\Temp\in\{825,865,895,915,935,950,960,975,1020\}^\circ\text{C}$. For each reduction-temperature pair, EBSD maps may be available in either RD or TD. However, not all $(\Roll,\Temp,\Dir)$ combinations are present in the dataset (see \autoref{fig:data_split}). In particular, for $\Roll=94\%$, data are available only at $\Temp\in\{825,915\}^\circ\text{C}$ and $\Dir=\TD$. 

The TexFlow model is then trained to learn the conditional distribution $p(\bfR|\Roll,\Temp,\Dir)$. To evaluate interpolation and extrapolation capabilities, selected combinations of $(\Roll,\Temp,\Dir)$ are simultaneously withheld from the training set, as indicated in \autoref{fig:data_split}. For each condition, whether in training or testing set, we randomly sample $N=100{,}000$ rotations for input to TexFlow.

\textbf{In-distribution performance}

\begin{figure}
    \centering
    \includegraphics[width=\textwidth]{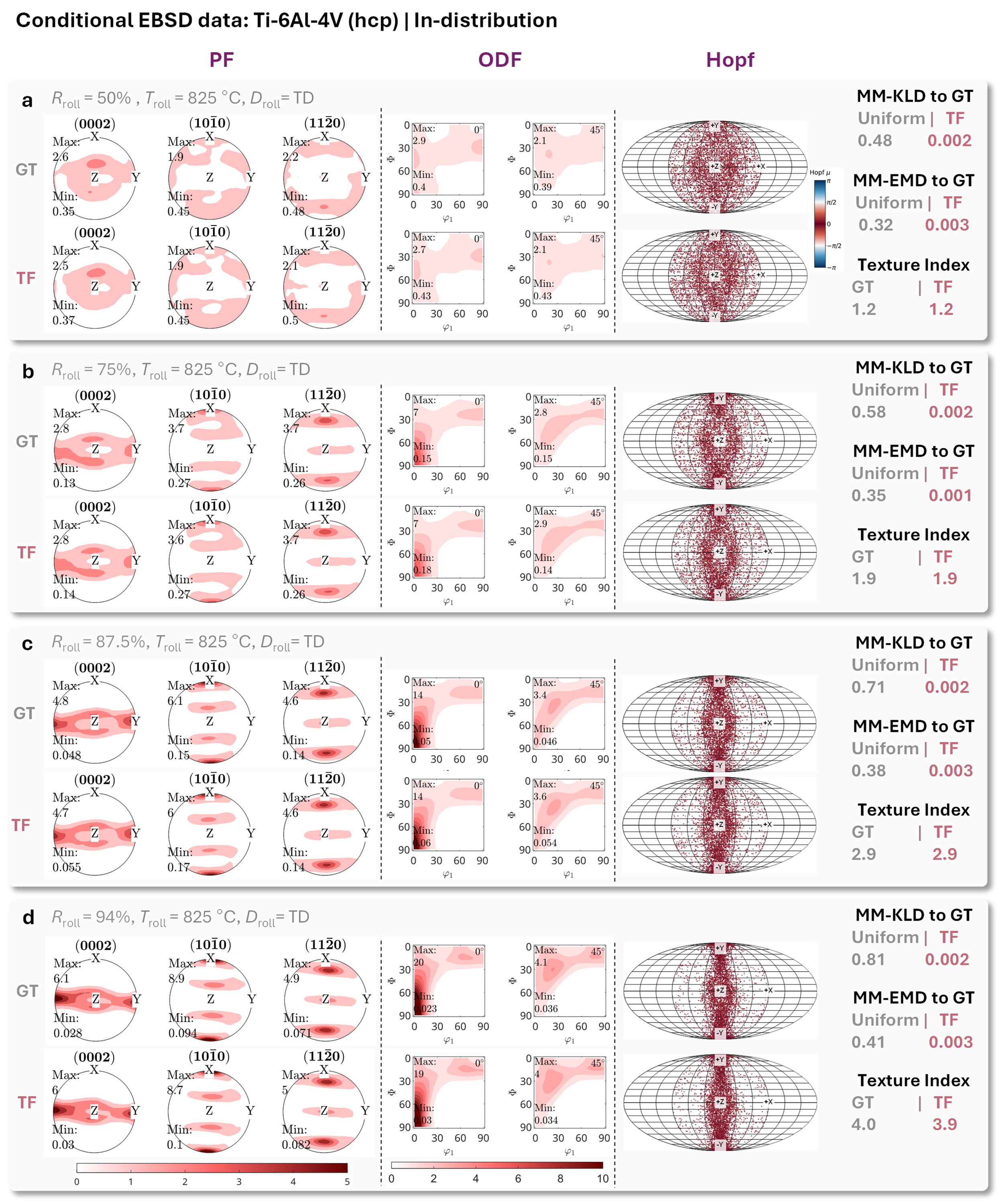}
    \caption{
    \textbf{Comparison of TexFlow vs. ground truth on conditional EBSD data of Ti-6Al-4V (hcp) on seen data:} at condition $\Temp = 825^\circ\text{C}$, $\Dir = \TD$ and variable rolling reduction: (a) $\Roll=50\%$,(b) $\Roll=75\%$,(c) $\Roll=87.5\%$,(d) $\Roll=94\%$ . Each panel shows (from left-to-right) pole figures (PF), ODF sections, Hopf plots, and quantitative metrics: MM-KLD, MM-EMD, and texture index.
    }
    \label{fig:cond_ti64_ebsd_in_dist}
\end{figure}

The performance of the model on in-distribution conditions (i.e., those included in the training set)  is presented in \autoref{fig:cond_ti64_ebsd_in_dist}. The qualitative comparisons based on pole figures, ODF sections and Hopf plots show that the model captures the progressive strengthening of texture with increasing $\Roll$. This trend is particularly evident at $825^{\circ}C$, where TexFlow reproduces the strengthening of rolling texture components, including the transverse basal component $\{0002\}\parallel\TD$ and the prismatic component $\left\{11\bar{2}0\right\}\langle 10\bar{1}0\rangle$ in agreement with the experimentally observed texture evolution \cite{Zeng2025EffectRolling}. 

\autoref{fig:cond_ti64_ebsd_in_dist} also provides a quantitative comparison of TF and GT distributions using MM-KLD, MM-EMD and texture index values.  The MM-KLD and MM-EMD of a uniformly random texture relative to the GT data serve as reference values representative of a model that fails to learn the underlying texture. At the observed conditions, the MM-KLD values for TF distribution are close to $0.002$ while the MM-EMD values range from $0.001$ to $0.003$. These values are significantly lower than the reference values corresponding to the uniformly random texture, indicating positive agreement between TF and GT. The texture index also increases with increasing $\Roll$ value, from $1.2$ at $50\%$ to $4.0$ at $94\%$; TF predictions follow this trend and the texture index values are close to the GT values. Taken together, these results demonstrate that TexFlow successfully learns the relationship between rolling reduction and texture evolution.

\textbf{Generalizability: Interpolation}

\begin{figure}
    \centering
    \includegraphics[width=\textwidth]{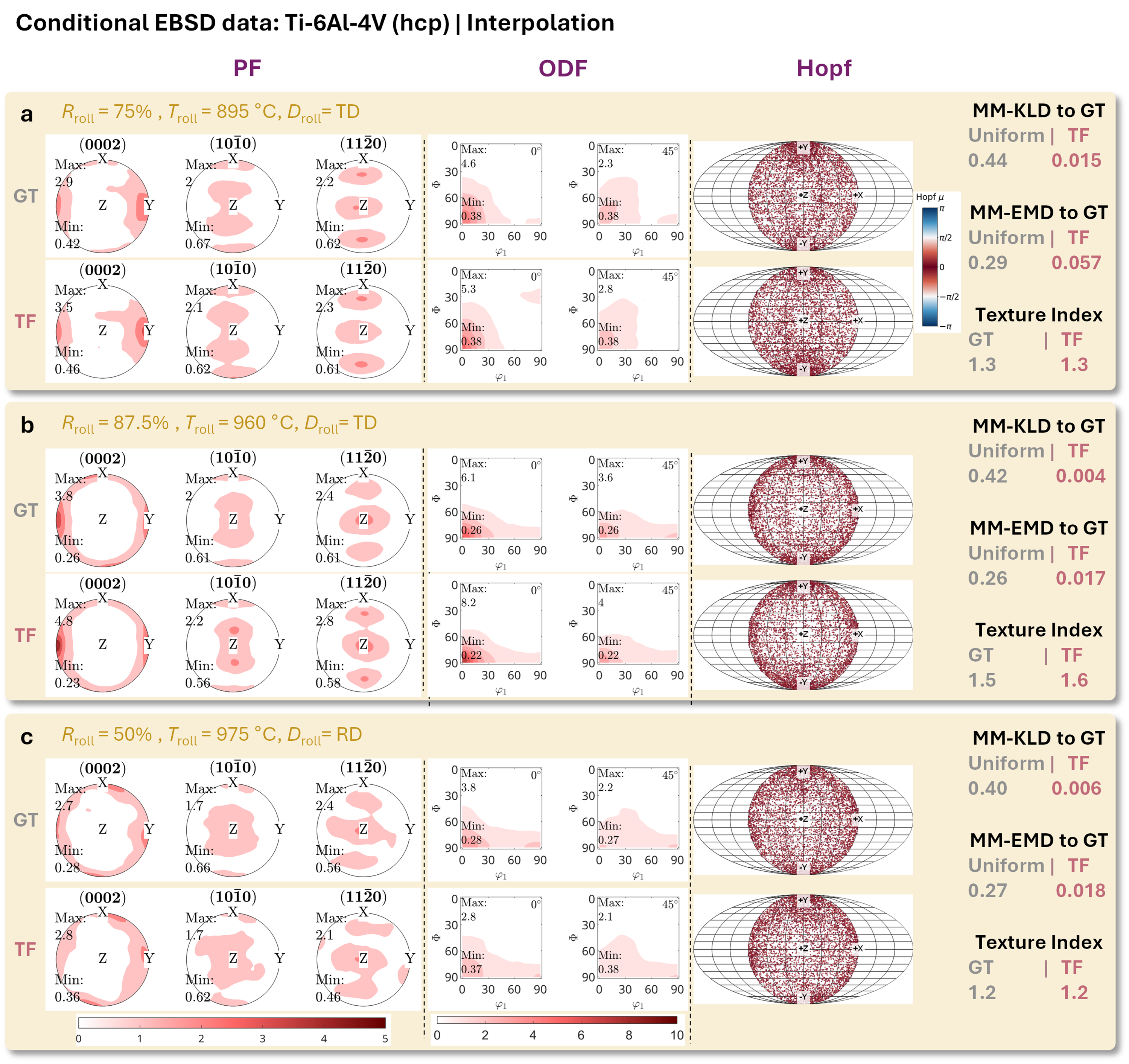}
    \caption{
    \textbf{Comparison of TexFlow vs. ground truth on conditional EBSD data of Ti-6Al-4V (hcp) for interpolation:} (a,c,d) observed conditions and (b) unseen condition for interpolation. Each panel shows (from left-to-right) pole figures (PF), ODF sections, Hopf plots, and quantitative metrics: MM-KLD, MM-EMD, and texture index. 
    }
    \label{fig:cond_ti64_ebsd_interp}
\end{figure}

\autoref{fig:cond_ti64_ebsd_interp} compares the TF and GT textures at unseen conditions $(\Roll,\Temp,\Dir)$: $(75\%,895^\circ\text{C}, \TD)$,
$(87.5\%, \allowbreak 960^\circ\text{C}, \allowbreak  \TD)$, and
$(50\%,975^\circ\text{C}, \RD)$,
which were deliberately withheld from the training set to evaluate interpolation performance. The pole figures, ODF sections, and Hopf maps indicate good agreement between TF and GT texture across all reductions, with the TF samples reproducing the principal clusters and overall spread observed in the GT data. 

TexFlow successfully captures the evolution of the dominant $\alpha$-phase rolling texture. At the condition  $(75\%, \allowbreak 895^\circ\text{C}, \allowbreak  \TD)$, the texture is characterized by a transverse basal alignment,  $\{0002\} \parallel \TD$, with the typical hot rolling prismatic component, $\{11\bar{2}0\}\langle 10\bar{1}0\rangle$. At $(87.5\%,960^\circ\text{C}, \TD)$, the prismatic component persists, whereas the transverse basal alignment weakens relative to its maximum at intermediate temperatures. At $(50\%,975^\circ\text{C}, \RD)$, the $\alpha$-phase texture is comparatively weak and diffuse, with an increasing $(0002) \parallel \RD$ contribution near the $\beta$-transus temperature \cite{Zeng2025EffectRolling}.

 Across the different conditions, the MM-KLD and MM-EMD values are in the range of $0.004$ and $0.06$, respectively, approximately an order of magnitude larger than those at the observed conditions, as expected for an unseen condition. Nevertheless, both values remain substantially below the corresponding reference values for a uniformly random texture, indicating good interpolation accuracy. The predicted texture index values of TF samples are in good agreement with those of the GT, while the predicted distribution remains in good agreement with the GT in both qualitative and quantitative comparisons. Overall, these results demonstrate that TexFlow can accurately interpolate the conditional orientation distribution between observed processing conditions.

\begin{figure}
\centering
\includegraphics[width=\textwidth]{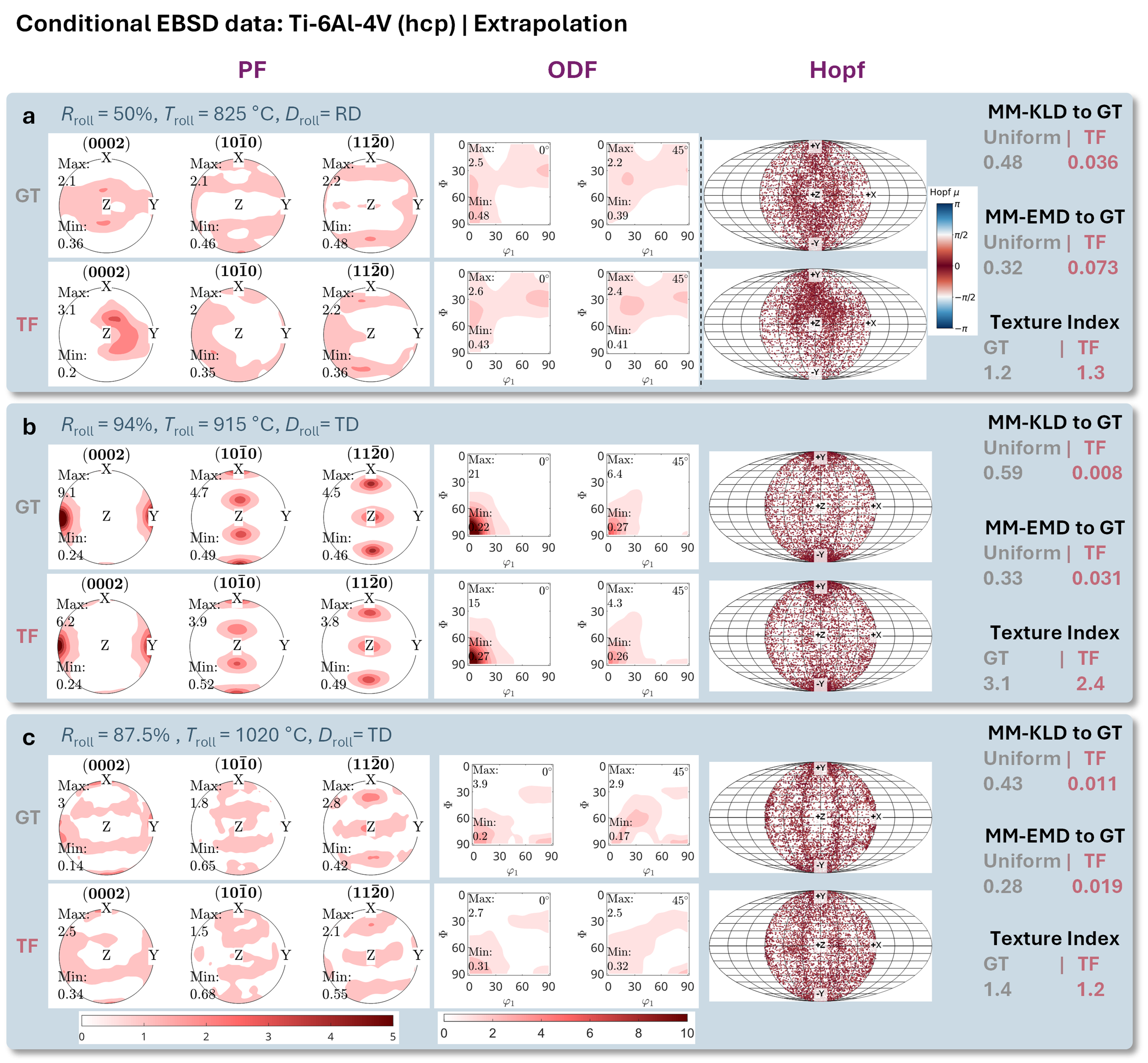}
 \caption{\textbf{Comparison of TexFlow vs. ground truth on conditional EBSD data of Ti-6Al-4V (hcp) for extrapolation:} (a,b,c) observed conditions and (d) unseen condition for extrapolation. Each panel shows (from left-to-right) pole figures (PF), ODF sections, Hopf plots, and quantitative metrics: MM-KLD, MM-EMD, and texture index.
}
\label{fig:cond_ti64_ebsd_extrap}
\end{figure}

\textbf{Generalizability: extrapolation}

\autoref{fig:cond_ti64_ebsd_extrap} compares the TF and GT textures at three unseen conditions  $(\Roll,\Temp,\Dir)$: $(50\%, 825^\circ\text{C}, \RD)$,
$(94\%, 915^\circ\text{C}, \TD)$, and
$(87.5\%, 1020^\circ\text{C}, \TD)$ , which were also withheld from the training set to evaluate extrapolation performance. The qualitative plots such as pole figures, ODF sections, and Hopf maps show reasonable agreement between TF and GT texture at some conditions. 

For the $(50\%, 825^\circ\text{C}, \RD)$ case (\autoref{fig:cond_ti64_ebsd_extrap}a), the pole figures show noticeable differences in the location and shape of intensity maxima relative to ground truth. The ODF sections reproduce some of the diffuse features of the reference distribution, while the Hopf plots show a broadly similar distribution. The texture index is closely reproduced, with values of $1.3$ and $1.2$ for the TF and GT distributions, respectively. Overall, the texture was not recovered for this particular condition.

A closer qualitative agreement is observed for the $(94\%, 915^\circ\text{C}, \TD)$ case (\autoref{fig:cond_ti64_ebsd_extrap}b). TexFlow captures the dominant features of the texture as seen in the pole figures and the ODF sections showing the basal and prismatic components. The Hopf plots also show a similar rotation distribution. The MM-KLD and MM-EMD values are lowest among the three cases considered for extrapolation, namely $0.008$ and $0.031$, respectively, compared with the references of a uniformly random distribution of rotation. The predicted texture index $2.4$ is lower than the GT value of $3.1$, indicating the predicted texture being weaker than the ground truth data.

Finally, for the  $(87.5\%,1020^\circ\text{C},\TD)$ case (\autoref{fig:cond_ti64_ebsd_extrap}c), the ground truth texture is relatively weak and diffuse. TexFlow reproduces the overall texture as seen in the pole figures, ODF sections and Hopf plots, although the predicted peak intensities and component shapes differ slightly from the ground truth. The quantitative agreement is strong, with MM-KLD and MM-EMD values of $0.011$ and $0.019$ respectively. The texture index of predicted texture $1.2$ is close to the ground truth value of $1.4$.

Overall, the results demonstrate that TexFlow can reproduce the texture under extrapolation. However, the pole figures and ODF sections reveal that the model does not consistently predict the texture at the edge case of unseen conditions. The extrapolation results demonstrate partial generalization, where TexFlow captures overall texture character, but the component level accuracy decreases when predicting texture at conditions outside the training distribution. The reduced extrapolation accuracy at the unseen condition is expected and can be attributed to the sparsity of the training data, with only 56 processing conditions available for training along three axes of conditions as seen in \autoref{fig:data_split}.

\subsection{Case 3: Conditional generation on CPFEM data} 
\label{subsec:res_condNF_cpfem}

We simulate polycrystalline copper using CPFEM to evaluate the performance of the conditional TexFlow model in learning texture evolution. The dataset is generated using the PRISMS-Plasticity framework \citep{Yaghoobi2022PRISMS-PlasticityPipeline}. The simulations model  the deformation of a face-centered cubic (FCC) copper polycrystalline microstructure under several loading paths, providing time-resolved orientation data suitable for learning process-texture relationships. 

\textbf{Numerical setup:}
The material system considered is polycrystalline copper with cubic symmetry, representative of Oxygen Free Copper (OFHC) commonly used in CPFEM studies \citep{Bronkhorst1992PolycrystallineMetals,Anand1996APlasticity,Lee1996TextureCopper}. The initial microstructure is represented by a representative volume element (RVE) with 197 grains on a $64\times64\times64$ cubic voxel grid. The RVE is discretized using a $16\times16\times16$ mesh of first-order hexahedral elements. Each element uses $2\times2\times2$ quadrature points, giving $32{,}728$ integration points. Periodic boundary conditions are imposed on the RVE, following standard CPFEM practice. 
Slip is assumed to occur on the twelve $\{111\}\langle 110\rangle$ FCC slip systems.

The constitutive parameters are adopted from established crystal plasticity studies \citep{Yaghoobi2022PRISMS-PlasticityPipeline}. Further implementation details of CPFEM are provided in the  PRISMS-Plasticity paper and documentation \citep{Yaghoobi2022PRISMS-PlasticityPipeline}.

\textbf{Dataset:}
Four different deformation modes are considered: uniaxial compression (C), uniaxial tension (T), pure shear (PS), and simple shear (SS), represented by the average deformation gradients imposed periodically on the boundaries:
\be
\label{def_grad}
    \bfF_{\mathrm{C}} = 
    \begin{bmatrix}
        1-0.5\hat t & 0 & 0 \\
        0 & 1 & 0 \\
        0 & 0 & 1
    \end{bmatrix},\ 
    \bfF_{\mathrm{T}} = 
    \begin{bmatrix}
        1+0.5 \hat t & 0 & 0 \\
        0 & 1 & 0 \\
        0 & 0 & 1
    \end{bmatrix},\ 
    \bfF_{\mathrm{SS}} = 
    \begin{bmatrix}
        1 & 0 & 0.5 \hat t\\
        0 & 1 & 0 \\
        0 & 0 & 1
    \end{bmatrix},\ 
    \bfF_{\mathrm{PS}} = 
    \begin{bmatrix}
        1+0.5\hat t & 0 & 0 \\
        0 & 1 & 0 \\
        0 & 0 & (1+0.5\hat t)^{-1}
    \end{bmatrix},
\ee

with $\hat t= t/t_\text{max}$, which denotes normalized simulation time. The simulations are performed over $5{,}000$ increments with a constant time increment of  $\Delta t=0.01$, giving a total simulation time of $t_\text{max}=50$. These deformation modes provide a robust test, as they induce diverse, pronounced, and path-dependent texture evolution.  Simulation snapshots are saved every 100 increments, corresponding to an interval of one simulation time unit and yielding 50 snapshots per loading path. Each snapshot of the RVE contains $32{,}768$ orientations based on the number of the integration points. For each snapshot, $N=10{,}000$ orientations are randomly subsampled.  For each deformation mode, the texture data at each time increment is then paired with the corresponding time $t$ as the conditioning variable.

\textbf{In-distribution performance}

\begin{figure}
    \centering
    \includegraphics[width=0.9\textwidth]{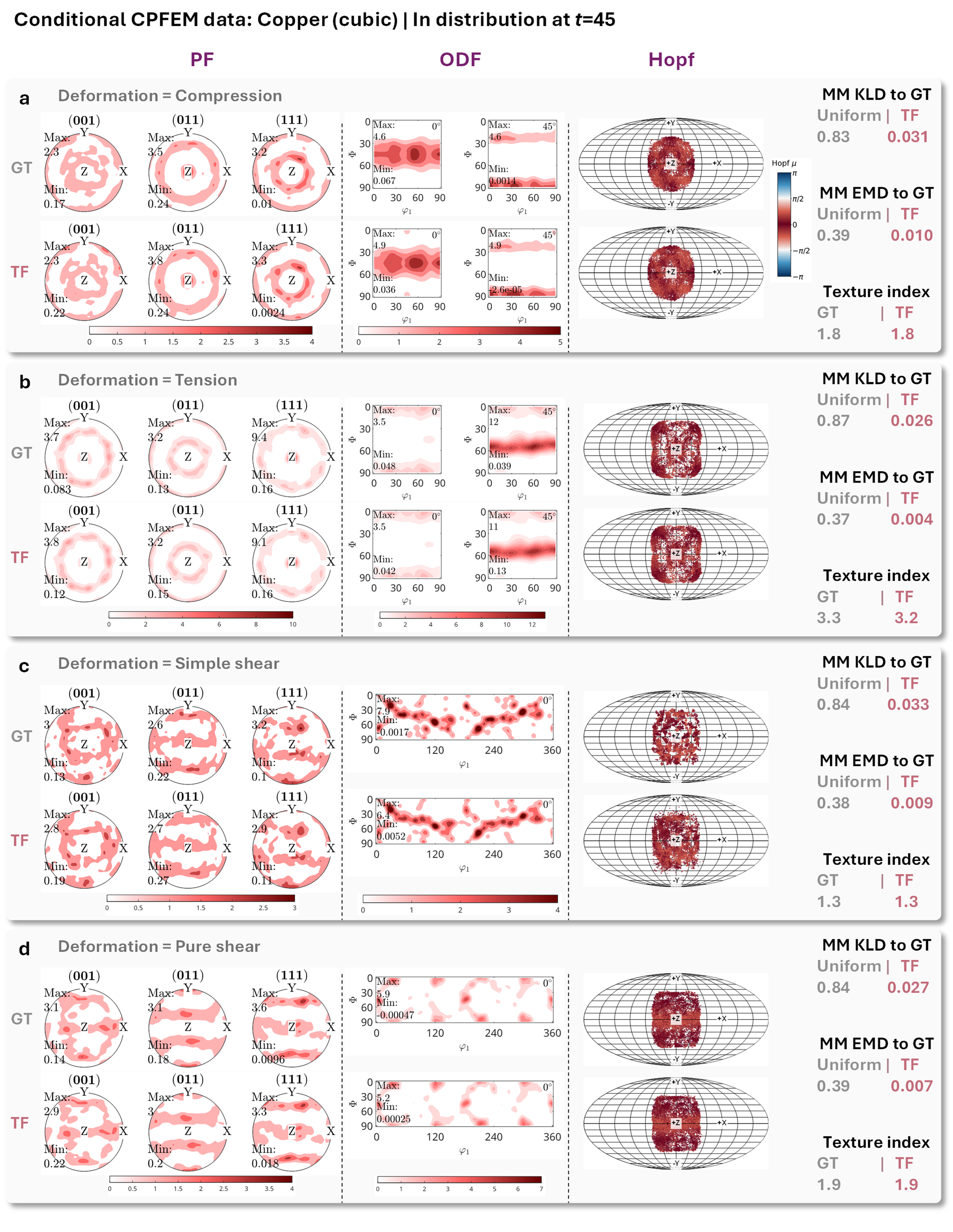}
    \caption{\textbf{Comparison of TexFlow vs.~ground truth on conditional CPFEM data of copper (cubic) for the in-distribution condition} at $t=45$ in  four deformation modes: (a)  compression, (b) tension, (c) simple shear, and (d) pure shear. Each panel shows (from left-to-right) pole figures (PF), ODF sections, Hopf plots, and quantitative metrics: MM-KLD, MM-EMD, and texture index.
    }
    \label{fig:def_mode_cpfem}
\end{figure}

The qualitative and quantitative results at condition $t=45$ are presented in \autoref{fig:def_mode_cpfem}. For each deformation mode, the pole figures, ODF sections and Hopf maps compare the texture components predicted by TexFlow with the corresponding GT data. The right-hand panels report the MM-KLD, MM-EMD and texture index values for a comprehensive quantitative assessment.

\ul{Uniaxial compression (C) and tension (T):} For uniaxial loading, the pole figures and ODF sections in \autoref{fig:def_mode_cpfem}(a,b) show the development of progressively strengthened axisymmetric texture fibers aligned with the loading direction. Under compression, $\langle001\rangle$ and $\langle011\rangle$ orientation families are prominent, consistent with classical CPFEM and experimental observations of FCC copper under compressive loading  \citep{Sato2022CrystallographicPeening,Yaghoobi2022PRISMS-PlasticityPipeline,Lee1996TextureCopper}. TexFlow reproduces the corresponding intensity distributions in the pole figures as well as the ODF features. 

For tensile deformation, the texture evolves toward a dominant $\langle111\rangle$ fiber aligned with the tensile axis, becoming increasingly pronounced with time, which is in agreement with earlier studies on Copper \citep{Wang2021AnTension, Kalidindi1992AnMetals}. TexFlow reproduces this fiber texture and its gradual sharpening with time, indicating the conditional model captures the evolution of rotation distributions toward stable components.

The quantitative metrics support these observations. For compression, TexFlow gives low MM-KLD of $0.031$, low MM-EMD  of $0.01$, and a texture index of $1.8$ (which matches with GT value). For tension, the corresponding values are $0.026$, $0.004$ and $3.2$ (compared with GT value of $3.3$), respectively as seen in the right panels of \autoref{fig:def_mode_cpfem}. 
These results indicate the model captures the orientation distribution and the increase in texture strength with deformation types of compression and tension. 

\ul{Simple shear (SS) and pure shear (PS):} Simple shear exhibits the most complex texture evolution among the considered loading paths. The texture under simple shear remains highly heterogeneous, with persistent multimodality and limited collapse toward a single fiber.  TexFlow reproduces these features  in \autoref{fig:def_mode_cpfem}(c). The corresponding pole figures and ODF sections reveal strongly non-axisymmetric distributions associated with the development of the classic FCC shear texture components  ($\mathrm{A}$, $\bar{\mathrm{A}}$, $\mathrm{B}$, $\bar{\mathrm{B}}$, and $\mathrm{C}$), arising from sustained lattice rotation \citep{Suwas2019TextureProcesses,Zhang2020EffectCopper,Azzeddine2022TextureProcessing}.  The quantitative comparisons give a low MM-KLD of $0.033$, a low MM-EMD of $0.009$, and a texture index of $1.3$ matching that of GT.  

Similarly, for pure shear, TexFlow captures the multi-component nature of the GT texture and reproduces the principal intensity variations in the pole figures and ODF sections as seen in \autoref{fig:def_mode_cpfem}(d). The Hopf maps show comparable cluster geometry and orientation spread for the two distributions. Quantitatively, the TF prediction gives a low MM-KLD of $0.027$ and low MM-EMD of $0.007$, and a texture index of $1.9$ matching with the GT value. These close agreements demonstrate that the model can represent a relatively diffuse orientation distribution associated with shear deformations.
 
\textbf{Generalizability: interpolation and extrapolation  }

\begin{figure}
    \centering
    \includegraphics[width=\textwidth]{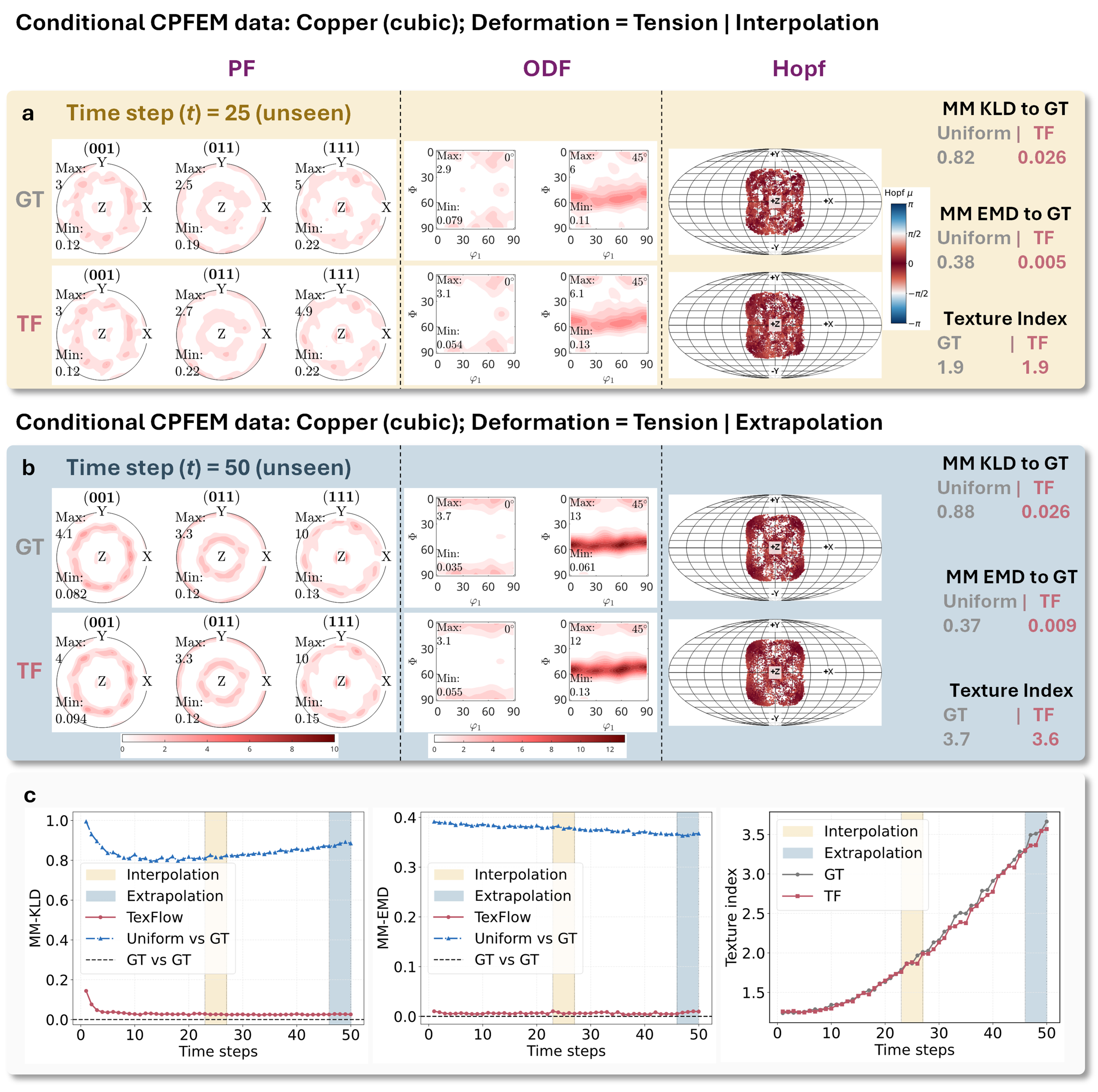}
    \caption{\textbf{Comparison of TexFlow vs.~ground truth on conditional CPFEM data of copper (cubic) for interpolation and extrapolation} at $t=25$ and $t=50$, respectively, in uniaxial tension deformation mode. Each panel shows (from left-to-right) pole figures (PF), ODF sections, Hopf plots, and quantitative metrics: MM-KLD, MM-EMD, and texture index.
    }
    \label{fig:cpfem_extra_interp}
\end{figure}

The generalizability of TexFlow is further assessed using the CPFEM dataset with the deformation mode being uniaxial tension. TexFlow is simultaneously evaluated in two settings:
\begin{itemize}
    \item interpolation (\autoref{fig:cpfem_extra_interp}(a)) at condition $t=25$, with data for  $t=(23 ,\ldots,27)$ excluded from training set,
    \item extrapolation (\autoref{fig:cpfem_extra_interp}(b)) at condition $t=50$, with data for $t=(46,\ldots,50)$ excluded from training set. 
\end{itemize} 

At the interpolation condition $t=25$, TexFlow  reproduces the dominant tensile texture components observed in the GT data. The pole figures and ODF sections show the strengthening of the $\langle111\rangle$ fiber associated with uniaxial tension. The Hopf maps also reproduce the location and overall spread of the rotation cluster. The quantitative metrics support the above agreement, where the TF prediction gives a low MM-KLD of $0.026$, a low MM-EMD of $0.005$ and a texture index of $1.9$ matching that of GT.

The extrapolation result at $t=50$ remains qualitatively and quantitatively consistent with the GT distribution. TexFlow captures the continued strengthening of texture components and reproduces the main features in pole figures, ODF sections and Hopf maps. MM-KLD and MM-EMD remain low at $0.026$ and $0.009$. The model slightly underestimates the final texture index at $3.6$, compared to $3.7$ of GT. 

The evolution of the different metrics with respect to the condition $t$ is visualized in the \autoref{fig:cpfem_extra_interp}(c). The MM-KLD and MM-EMD computed between the TF and GT distributions remain close to zero at all times throughout the loading process and are substantially lower than those computed between uniformly random textures and the GT data (a baseline representing a model that fails to learn the underlying texture distribution).
The texture index progressively increases with $t$, reflecting the  strengthening and sharpening of $\langle111\rangle$ fiber texture component. TexFlow predictions follow this trend closely over the entire range of $t$ values, including the interpolation and extrapolation regime highlighted in the figure.

The close agreement at $t=25$ demonstrates that TexFlow can interpolate between observed temporal conditions while preserving the structure of the conditional orientation distribution. The result at $t=50$ shows that the model also provides reasonable extrapolation beyond the training range, although a small underestimation of the texture index becomes apparent at the furthest time increment. The CPFEM results show relatively better extrapolation performance compared to the conditional EBSD results (\autoref{subsec:res_condNF_ebsd}), which is partly due to the denser sampling of the conditioning variable. Time-resolved evolution of the texture from CPFEM and TF for each deformation mode is further visualized in Supplementary Movies ESM 1-4 (compression, tension, simple shear, and pure shear, respectively) and shows good agreement. Overall, the results demonstrate that TexFlow can serve as an efficient surrogate for texture evolution with the capability to interpolate or extrapolate missing texture data.

\subsection{Ambiguous edge case} 
\label{subsec:diagnostic}

Schaeben et al. \citep{Schaeben2025CommentRepresentation}, in their comment on the work of Wanni et al. \cite{Wanni2024MachineRepresentation}, argued that projection-based representations of texture, e.g., through pole figures, can be misleading and may incorrectly classify two distinct textures as identical. We use the example presented by Schaeben et al. \citep{Schaeben2025CommentRepresentation} as an edge case to demonstrate that such ambiguities do not arise in TexFlow, as it learns texture directly on rotations rather than through their projections.

\begin{figure}
    \centering
    \includegraphics[width=0.9\textwidth]{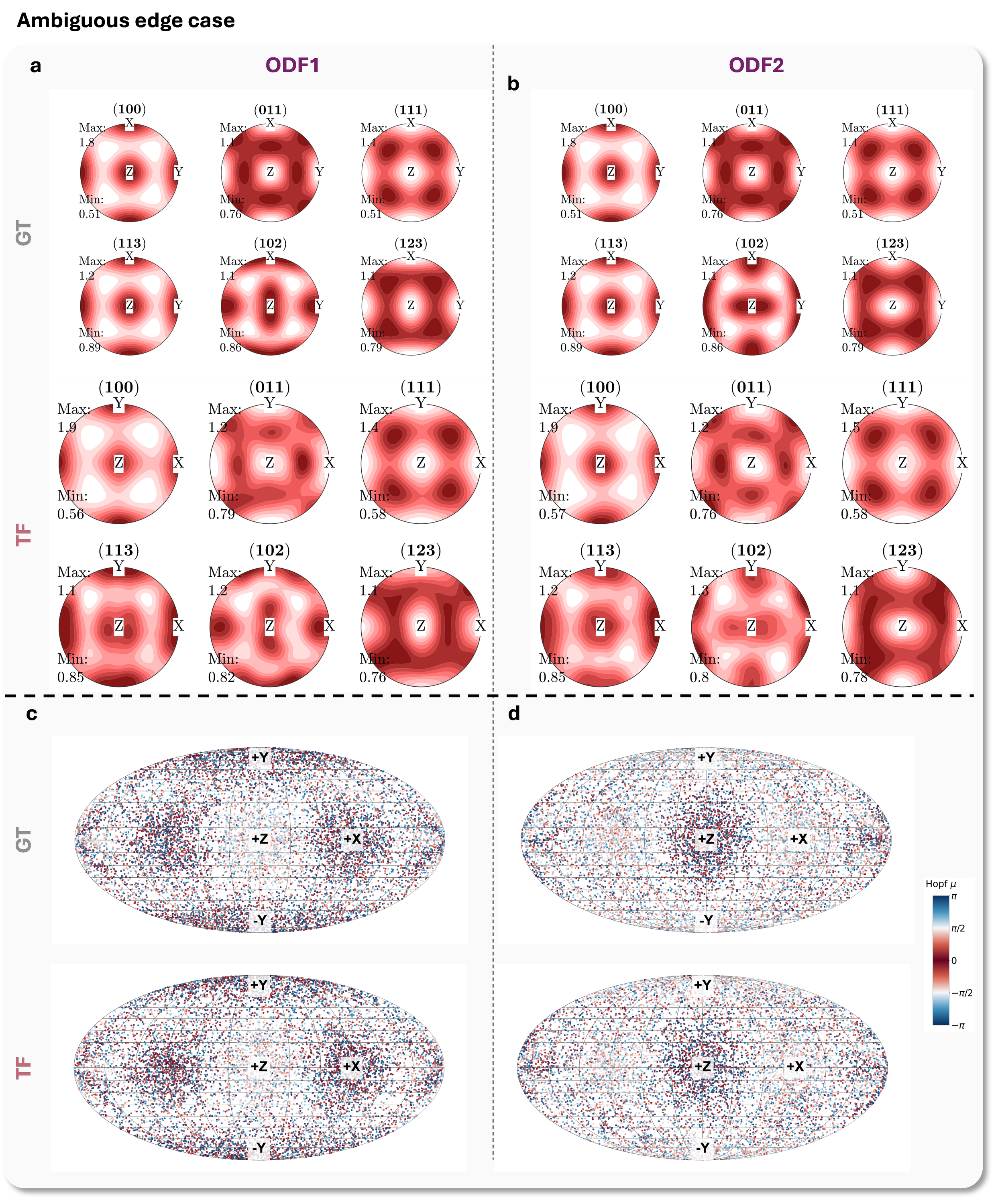}
    \caption{\textbf{Edge case illustrating projection ambiguity on $\SO$.}  Two distinct orientation distributions, ODF1 and ODF2, are compared using conventional pole figures and Hopf maps. Both ground truth (GT) and TexFlow-generated textures are shown. Panels (a) and (b) show the pole figures for planes $(100)$, $(011)$, $(111)$, $(113)$, $(102)$, and $(123)$ for ODF1 and ODF2, respectively. The differences are visible in higher order poles $(102)$ and $(123)$. Panels (c) and (d) show the Hopf maps for the ODF1 and ODF2, respectively. This example demonstrates that TexFlow, by learning distributions directly on $\SO$, can resolve ambiguities that would otherwise arise when learning from projections such as pole figures. 
    }
    \label{fig:diag_case}
\end{figure}

In this construction (\autoref{fig:diag_case}), two distinct orientation distribution functions, denoted ODF1 and ODF2, are defined on $\SO$ such that they generate nearly indistinguishable pole figures and matching Euler-angle marginals, despite fundamental differences in their joint orientation distributions. \autoref{fig:diag_case}(a-b) illustrates this phenomenon, where the first several pole figures are identical, and differences become apparent only for selected higher-order poles, particularly $(102)$ and $(123)$. This behavior is related to a classical result: if two ODFs differ only in their \textit{odd} Fourier coefficients, their lower-order pole figures match  \citep{Bachmann2010TextureToolbox}. This edge case highlights that agreement in pole figures or Euler angle marginals is a \emph{necessary but not sufficient} condition for distributional correctness on $\SO$.  In particular, the similarity of lower-order projections can hide differences in joint orientation distribution.

We train TexFlow independently on both these ODFs and summarize the results. The pole figures obtained from TexFlow-generated rotations (\autoref{fig:diag_case}(a-b)) closely agree with their respective GT pole figures, including those for the $(102)$ and $(123)$ poles. When the TexFlow-generated and GT textures are visualized using Hopf maps (\autoref{fig:diag_case}(c-d)), the differences between the two textures become evident, resolving the ambiguity. This provides evidence that the model captures the orientation distributions on $\SO$ beyond individual pole figures or marginal projections.

\section{Conclusion}
\label{sec:conclusion}

We introduce \textbf{TexFlow}, a conditional normalizing-flow framework for probabilistic modeling of crystallographic texture on the rotation group $\SO$. TexFlow combines neural spline flows with a six-dimensional Gram-Schmidt orthogonalization (GSO) representation, enabling learning in an unconstrained Euclidean space while ensuring that generated samples correspond to valid rotation matrices $\bfR\in\SO$. In addition, Hopf-fibration maps provide a compact, global visualization of orientation distributions, complementing conventional pole figures and ODF sections while mitigating ambiguities and limitations associated with such projections.

TexFlow was evaluated on three categories of texture data: experimentally measured EBSD data from additively manufactured SS316L, forsterite, and glaucophane; conditional EBSD data capturing texture evolution in hot-rolled Ti-6Al-4V; and time-resolved conditional CPFEM data for FCC copper subjected to different loading paths. Across these datasets, TexFlow successfully reproduced multimodal orientation distributions and, in conditional cases, their evolution with process conditions. Qualitative agreement with ground-truth distributions was demonstrated through pole figures, ODF sections, and Hopf maps, while quantitative agreement was supported by low MM-KLD and MM-EMD values and accurate predictions of the texture index. Interpolation and extrapolation studies further demonstrated the model's ability to predict texture at previously unseen conditions, including conditions outside the training range. Collectively, these results demonstrate the potential of TexFlow as a flexible and computationally efficient probabilistic framework for modeling crystallographic texture and its evolution.

Despite these promising results, several aspects of the TexFlow framework warrant further development:
\begin{itemize}
    \item \textbf{Likelihood formulation on $\SO$:} The current framework learns probability densities in the six-dimensional GSO coordinate space with respect to the Lebesgue measure, while generated samples are subsequently mapped onto $\SO$. Consequently, the learned probability densities do not directly correspond to probability densities with respect to the invariant Haar measure on $\SO$. Establishing such a correspondence requires accounting for the geometric distortion and many-to-one nature of the GSO mapping \cite{Man2022CrystallographicRepresentations}. Future work could develop a measure-consistent likelihood formulation, for example through an appropriate change-of-variables or co-area-based treatment, enabling likelihood evaluation directly on $\SO$ with respect to the Haar measure.
    \item \textbf{Explicit incorporation of crystallographic symmetry:} In the present implementation, orientations are mapped to the fundamental zone before training, and crystallographic symmetry is accounted for during post-hoc texture analysis. As a result, generated orientations are restricted to the fundamental zone rather than being modeled through an explicitly symmetry-aware probability distribution. Future extensions could incorporate crystallographic symmetry directly into the flow architecture, for example through symmetry-equivariant transformations or group-symmetrized probability densities, ensuring consistent treatment of symmetry-equivalent orientations by construction.
    \item \textbf{Incorporation of spatial microstructural information:} The current framework models orientation distributions without explicitly accounting for grain morphology, spatial correlations, or microstructural heterogeneity. Extending TexFlow to jointly model crystallographic orientations and spatial microstructural descriptors would enable a more comprehensive representation of microstructure and broaden its applicability in building process-structure-property maps.
\end{itemize}

Beyond these methodological improvements, TexFlow offers several opportunities for applications in materials characterization, processing, and modeling:
\begin{itemize}
    \item \textbf{Texture data augmentation and imputation:} The conditional generative capability of TexFlow could be used to estimate texture in regions of the process-texture-property space where experimental data are sparse or unavailable. This is particularly relevant for high-dimensional processing spaces, where comprehensive experimental characterization is costly and measurements may be affected by noise.
    \item \textbf{Construction of process-texture-property maps:} By coupling TexFlow with suitable texture-dependent property prediction models, probabilistic process-texture-property relationships could be constructed from limited datasets. Such maps could facilitate the identification of processing conditions that produce desired textures, as well as textures associated with targeted material properties.
    \item \textbf{Accelerating multiscale simulations:} Multiscale simulations often require computationally expensive mesoscale calculations (e.g., RVE-level CPFEM simulations) to capture microstructural evolution at each macroscale load increment and integration point. TexFlow could serve as a computationally efficient surrogate for mesoscale texture evolution by predicting orientation distributions that would otherwise require repeated RVE-level simulations, thereby reducing the overall computational cost.
\end{itemize}

\section{Data availability}
\label{sec:data_avail}
{ \textit{The datasets will be made available at the time of publication}}.

\section{Code availability}
\label{sec:code_avail}
{ \textit{The codes used in this work will be made available at the time of publication}}.

\section{Acknowledgments}
{This work was supported by the Dutch Research Council (NWO) through the NWO Veni Talent Program, project number 20214.}

\newpage

\appendix

\section{Architecture and training}
\label{app:arch_train}

This appendix summarizes the architecture and training settings used for the unconditional and conditional TexFlow models. All experiments employ the same six-dimensional GSO representation and rational quadratic rolling-spline coupling (RQ-RSC) transformations. The base distribution is a centered beta distribution, and the spline transformations use $K=16$ bins over the interval $[-B,B]$, with $B=4$. The number of coupling layers, conditioner network width, learning rate and number of training epochs are adjusted according to the complexity of each dataset and whether the model is unconditional or conditional. 

The conditioner networks are implemented as fully connected Dense-ReLU multilayer perceptrons with two hidden layers. For the conditional models, the process variables are supplied to the conditioner network at each coupling layer, while the flow transforms only the orientation parametrization. The models are trained using the Adam optimizer with a batch size of $256$. The complete configuration for each experiment is reported in \autoref{tab:training_hyper}.

\begin{table}[htbp]
\centering
\caption{Summary of model architecture and training hyperparameters used in TexFlow}
\label{tab:training_hyper}
\renewcommand{\arraystretch}{1.25}
\begin{tabular}{p{3cm} p{1.5cm} p{1.6cm}  p{1.7cm}  p{1.6cm}  p{1.6cm}  p{1.6cm}}
\hline
\textbf{Parameter} & \textbf{Symbols} & Uncond 1$^{\mathrm{a}}$ & Uncond 2$^{\mathrm{b}}$ & Uncond 3$^{\mathrm{c}}$ & Cond 1$^{\mathrm{d}}$ & Cond 2$^{\mathrm{e}}$  \\
\hline

Base distribution$^{\mathrm{f}}$           & $ p_z(\bfz)$ & (\texttt{CB})   & (\texttt{CB}) &  (\texttt{CB})  & (\texttt{CB}) &  (\texttt{CB})  \\
    Spline bins count       & $K$           &  16       &   16  &   16   &   16   &   16     \\
Spline domain parameter     &  $B$          &  4.0      & 4.0    & 4.0  & 4.0 & 4.0  \\
Spline domain               &   $[-B,B]$    &   $[-4,4]$    &   $[-4,4]$    & $[-4,4]$ & $[-4,4]$ & $[-4,4]$  \\
Number of coupling layers   &  $L$          &  20 &  9  &  18  &   18    & 16           \\ 
Coupling structure          & - & RQRSC & RQRSC & RQRSC & RQRSC & RQRSC \\
Conditioner network$^{\mathrm{g}}$         &    & 128  & 128  & 128  &  512 &  512 \\
Learning rate               & $\eta$  & $10^{-3}$  &  $10^{-3}$  &  $10^{-3}$ & $10^{-4}$  &  $10^{-4}$ \\
Optimizer                   &  $\mathcal{O}$ & Adam  & Adam  &  Adam & Adam  & Adam  \\
Batch size                  &  $N_{\mathrm{batch}}$ &  256 & 256  &  256 & 256  & 256   \\
Training epochs             &  $N_{\mathrm{ep}}$ & 200  & 200  & 200  & 100  & 100   \\

\hline
\end{tabular}
\vspace{2mm}
\begin{minipage}{\textwidth}
\footnotesize
$^{\mathrm{a}}$ Unconditional AM SS316 L dataset \\
$^{\mathrm{b}}$ Unconditional Forsterite dataset \\
$^{\mathrm{c}}$ Unconditional Glaucophane dataset \\
$^{\mathrm{d}}$ Conditional EBSD dataset of Ti-6Al-4V \\
$^{\mathrm{e}}$ Conditional CPFEM dataset of Copper \\
$^{\mathrm{f}}$ CB denotes the centered beta base distribution. \\ 
$^{\mathrm{g}}$ The conditioner is a fully connected Dense-ReLU MLP with two hidden layers and the reported value is the number of neurons in each hidden layer. \\
\end{minipage}
\end{table}

\section{Computational resources}
\label{app:compute}

All TexFlow models were implemented in Python using \texttt{JAX}. Training and inference tasks were performed using an  NVIDIA RTX A6000 GPU with 48GB VRAM, an AMD Ryzen Threadripper PRO 5965WX CPU with 24 physical cores and 48 logical threads, and 256~GB of system memory. The reported (in \autoref{tab:texflow_specs}) runtime correspond to the hardware and software configuration used for the experiments and are provided to give an estimate of the computational cost of model training and sampling.

\begin{table}[H]
\centering
\caption{Configuration and runtime for TexFlow experiments}
\label{tab:texflow_specs}
\renewcommand{\arraystretch}{1.25}
\begin{tabular}{p{1.5cm} p{3cm}  p{2.5cm}  p{2cm}  p{2.2cm}  p{2.5cm} }
\hline
\textbf{Task} & \textbf{Dataset / Case} & \textbf{Model Type} & \textbf{Hardware} & \textbf{Training time(s)} & \textbf{Sampling time(s)} \\
\hline

\multirow{3}{*}{Training} 
& AM SS316L  &           & GPU                   & 110 & 0.2  \\
& Forsterite & Unconditional          & GPU       & 85 & 0.5 \\
& Glaucophane &           & GPU                    & 188 & 0.9\\
\hline

\hline
Training  & EBSD  & Conditional        & GPU       & 19030 & 1.3\\
Training  & CPFEM  & Conditional       & GPU       & 838 & 1.6\\

\hline
\end{tabular}
\end{table}

\newpage

\bibliographystyle{unsrtnat}
\bibliography{references}

\end{document}